\documentclass[structabstract]{aa}
\usepackage{epsf}
\usepackage{graphicx}
\usepackage{natbib}
\bibpunct{(}{)}{;}{a}{}{,}

\def\blfootnote{\xdef\@thefnmark{}\@footnotetext}

\newcommand{\teff}{\ensuremath{T_{\rm{eff}}}}
\newcommand{\logg}{\ensuremath{\log g}}
\newcommand{\logy}{\ensuremath{\log y}}
\newcommand{\lheh}{\ensuremath{\log \left(N_{\mathrm{He}}/N_{\mathrm{H}}\right)}}

\newcommand{\chk}{\ensuremath{\surd}}
\newcommand{\hirac}{{\sc hirac}}
\newcommand{\alfosc}{{\sc alfosc}}

\newcommand{\twin}{{\sc twin}}
\newcommand{\uves}{{\sc uves}}
\newcommand{\twomass}{{\sc 2mass}}
\newcommand{\bal}{\object{Balloon\,090100001}}
\newcommand{\msol}{M\ensuremath{_\odot}}
\newcommand{\mpg}{\ensuremath{m_{\rm{pg}}}}
\defcitealias{silvotti00}{{Paper~\sc{i}}}
\defcitealias{ostensen01a}{{Paper~\sc{ii}}}
\defcitealias{ostensen01b}{{Paper~\sc{iii}}}
\defcitealias{silvotti02a}{{Paper~\sc{iv}}}
\defcitealias{solheim04}{{Paper~\sc{v}}}
\defcitealias{solheim06}{{Paper~\sc{vi}}}
\defcitealias{oreiro07}{{Paper~\sc{vii}}}
\defcitealias{oreiro09}{{Paper~\sc{viii}}}
\renewcommand{\listofobjects}{}
\begin{document}

\title{A survey for pulsating subdwarf B stars\\
with the Nordic Optical Telescope}

\author{
   R.~H.~{\O}stensen \inst{1}
   \and
   R.~Oreiro \inst{1}
   \and
   J.--E.~Solheim \inst{2}
   \and
   U.~Heber \inst{3}
   \and
   R.~Silvotti \inst{4}
   \and
   J.~M.~Gonz\'alez-P\'erez\inst{5}
   \and
   A.~Ulla\inst{6}
   \and
   F.~P\'erez Hern\'andez\inst{5,7}
   \and
   C.~Rodr{\'{\i}}guez-L{\'o}pez\inst{6,8}
   \and
   J.~H.~Telting\inst{9}
}

\institute{
Instituut voor Sterrenkunde, K.~U.~Leuven, Celestijnenlaan 200D, 3001 Leuven, Belgium\\
   \email{roy@ster.kuleuven.be, raquel@ster.kuleuven.be}
\and
Institutt for Teoretisk Astrofysikk, Universitetet i Oslo,
   0212 Blindern-Oslo, Norway% \\ \email{j.e.solheim@astro.uio.no}
\and
Dr.~Remeis-Sternwarte, Astronomisches Institut der Univ.~Erlangen-N\"urnberg,
   96049 Bamberg, Germany% \\ \email{heber@sternwarte.uni-erlangen.de}
\and
INAF-Osservatorio Astronomico di Torino, Strada dell'Osservatorio 20, 10025 Pino Torinese, Italy
\and Instituto de Astrof\'isica de Canarias, 38200 La Laguna, Tenerife, Spain
\and Departemento F\'isica Aplicada, Universidade de Vigo, 36310 Vigo, Spain
\and Departamento de Astrof\'\i sica, Universidad de La Laguna, 38205 La Laguna, Tenerife, Spain
\and Laboratoire d'Astrophysique de Toulouse-Tarbes, Univ.~de Toulouse,
     14 av.\,Edouard Belin, Toulouse 31400, France
\and Nordic Optical Telescope, 38700 Santa Cruz de La Palma, Spain
}

\date{Received 15 October 2009 / Accepted 6 January 2010 }

\abstract
{A search programme for pulsating subdwarf B stars was conducted with the Nordic Optical
 Telescope on La Palma over 59 nights between 1999 and 2009. 
}
{The purpose of the programme was to significantly extend the number of
 rapidly pulsating sdB stars to better understand the
 properties of this new group of variable compact stars.
}
{Candidates were selected initially from the HS and HE surveys, but
 were supplemented with additional objects from other surveys.
 Short sequences of time-series photometry were made on the candidates to determine
 the presence of rapid pulsations.
}
{In total twenty new pulsators were found in this survey, most of which have already been
 published and some extensively studied. We present four new
 short period pulsators, bringing the total of such pulsators up to 49.
 We also give limits on pulsation amplitudes for 285 objects
 with no obvious periodic variations,\thanks{
The complete list with data on of 285 objects (Table~\ref{tbl:limits}) is
only available in electronic form at the CDS via anonymous ftp to
{\tt cdsarc.u-strasbg.fr (130.79.128.5)}
or via {\tt http://cdsweb.u-strasbg.fr/cgi-bin/qcat?J/A+A/???/???}}
 summarise the results of the
 survey, and provide improved physical parameters on the
 composite pulsators for which only preliminary estimates were published
 earlier. 
}
{}

\titlerunning{A Survey for Pulsating sdB Stars}
\authorrunning{\O stensen, Oreiro, Solheim et al.}

\maketitle

\section{Introduction}

Hot subluminous stars are considered to be extreme horizontal branch (EHB)
stars or closely related objects, with effective temperatures
\teff\,$\simeq$\,20\,--\,35\,kK.  The EHB models imply that they
are core helium burning objects with an extremely thin 
($M_{\mathrm{env}} \leq 0.02 M_\odot$) inert hydrogen dominated envelope
~\citep{heber86,saffer94}.
This structure prevents them from ascending the asymptotic giant branch
(AGB), and they must evolve instead towards higher temperatures and surface
gravities after their core helium is exhausted. Thus, an sdB star crosses the 
hotter sdO domain before reaching degeneracy and cooling as a normal white
dwarf star \citep{dorman93}.  However, important questions remain 
concerning the exact evolutionary paths and the appropriate timescales.

How they evolve to the EHB configuration is controversial. 
The problem is how the mass loss mechanism in the progenitor manages
to remove all but a tiny fraction of the hydrogen envelope at
about the same time as the core has attained the mass
($\sim$\,0.47\,M$_\odot$) required for the He flash. 
About half of the sdB stars reside in close binary systems, with either
a white dwarf or an M-dwarf as companion
\citep{maxted01,napiwotzki05}, and a significant
fraction of the rest are in wider binaries with a main sequence F--K star
as companion. Therefore mass transfer in close binary evolution must be
an important evolutionary pathway leading to the formation of sdB stars
\citep{han02,han03}. A recent review of hot subdwarf stars is provided by
\citet{heber09}.

The discovery of multimode pulsations in sdBs opened an attractive opportunity
of probing their interiors with seismological methods.
The properties of the sdB pulsators
~\citep[sdBVs or \object{V361\,Hya}\footnote{
Translations between the old survey names and the variable star
names used here are provided in Table~\ref{tbl:sdbvars}.}
stars after the prototype;][]{kilkenny97} 
are characterised by relatively short pulsation periods
($\sim$1 to 10\,min) and low pulsation amplitudes attributed to
low order pressure ($p$) modes.
Most V361\,Hya stars have been found with pulsation amplitudes of around
ten millimodulation amplitudes (mma\footnote{The mma units are $10^{-3}$
of the Fourier amplitudes of a light-curve in normalised intensity units.
One mma translates to a peak-to-peak amplitude of two
millimodulation intensity (mmi) units in the light curve.}),
although a few objects show a main peak amplitude up to $\sim$60\,mma 
(\object{V338\,Ser}, \citealt{kilkenny99};~\bal, \citealt{oreiro04}),
while the lowest level pulsator discovered up to now is
\object{LM\,Dra}
\citep{silvotti00}, with no peaks above 2\,mma.
The number of detected periods varies from a single one to more than fifty,
strongly correlated with the accuracy of the measurements.

In 2001, some sdBs were discovered to show long-period ($\sim$1\,h)
photometric modulations \citep{green03}, which were interpreted as high
radial order gravity ($g$) modes. These stars are referred to as long period sdBV
stars or V1093\,Her stars after the prototype.
V1093\,Her stars are cooler than their short-period counterparts, although
their instability regions appear to overlap around
$T_{\rm eff}\sim$29\,kK, where they can display simultaneous short
and long period pulsations. These stars are referred to as hybrid sdBVs
or DW\,Lyn stars after the prototype \citep{schuh06}.

Non-adiabatic computations by~\cite{charpi97} predicted excited low
degree ($\ell$), low radial order ($n$) $p$-modes in sdB models,
driven by an iron opacity bump, in the temperature range 29\,--\,37\,kK. 
This mechanism is inefficient at solar metallicity,
but gravitational settling and radiative levitation can work together to
locally enhance metals in a driving zone in the envelope.
This $\kappa$ mechanism has been successfully invoked to explain both
the $p$-mode pulsations in V361\,Hya stars and the $g$-mode
pulsations in V1093\,Her stars \citep{fontaine03}.
\citet{jeffery06} were able to expand the instability regions
by including opacities for nickel, to the point where the
$p$-mode and $g$-mode domains overlap, thereby explaining
the presence of hybrid oscillations.
However, it is still not understood why most stars in this 
temperature range do not appear to vary. 

As the number of known short-period and hybrid sdBV stars have increased,
several peculiar features are emerging:
\begin{itemize}
\item {\em Mode density:}
The number of pulsation frequencies detected vary from a single one to more
than 50. 
\object{LS Dra} \citep{ostensen01b} was confirmed to be monoperiodic
well below the millimagnitude level from 127 hours of observations by
\citet{reed07a}.  For others, detailed follow-up observations have increased
the number of detected frequencies to more than 50.
\item {\em Period grouping:} Pulsation periods for the whole V361\,Hya group
lie between 90 and 800\,s, but distinctions appear to be present between
the stars in the shorter and longer period groups.
All the sdBVs with periods between 300 and 400\,s except one
are DW\,Lyn type hybrid pulsators.
The odd one out is \object{KL\,UMa}, which also stands out as the only
binary of the pulsators with periods in this range \citep{otoole04}.
\item {\em Amplitude variations:}
Amplitudes are found to change with time in many cases.
However, it is unclear whether this is due to true amplitude variations
or beating of closely spaced modes \citep{kilkenny10}.
\item {\em Dominant modes:}
According to their amplitudes the stars can be grouped into high
amplitude pulsators, for which a mode was observed to exceed
30\,mmag at least once, and low-amplitude ones for which such dominant
modes were never reported. All the high amplitude pulsators are DW\,Lyn
type, except the unique \object{V338\,Ser}, which lies well above the
canonical EHB region.
\end{itemize}

The V361\,Hya stars with very few exited modes are not suitable for the
period-matching technique in asteroseismology, since too many models
can be found that adequately match their sparse periodograms.
However, if a mode has sufficient amplitude, the degree $\ell$
can be determined from spectroscopy, substantially constraining
the asteroseismic solution.
For low-amplitude V361\,Hya stars with sufficiently rich spectra, however, 
asteroseismology has met with great success, as the most
fundamental parameter of a star -- its mass --
has been derived for ten such pulsators along with their fractional hydrogen 
envelope masses (see \citealt{ostensen09} for a review, and \citealt{randall09b}
for the most recent result).

Nevertheless, sdBVs with sparse pulsation spectra are important as they
may allow the measurement of light travel time variations or period changes
due to stellar evolution, if the modes are sufficiently stable, as exemplified
by the planet-hosting \object{V391\,Peg} \citep{silvotti07},
discovered by our survey \citep{ostensen01a}.
Hence, photometric surveys to discover new V361\,Hya stars are still rewarding.

The theoretical location of the sdBV instability strip in the \teff--\logg\
plane is quite well established.
However, no more than one out of ten sdB stars located within the instability 
region are actually found to pulsate. This situation is quite different 
from that of white dwarf pulsators of the ZZ Ceti type, for which evidence is
growing that all white dwarfs in their instability strip are pulsating
\citep{gianninas07}.
For sdBVs it has been suggested that younger EHB stars may not have
enough iron accumulated in the driving region for the $\kappa$-mechanism
to be efficient enough to drive pulsations, but 
time-dependent diffusion calculations by \cite{fontaine06} have
demonstrated that sufficient iron for pulsations to occur accumulates after
a few hundred thousand years, and after 1\,Myr no further accumulation
is achieved. Compared to the EHB lifetime of 100\,--\,150\,Myr this is 
not significant. An interesting speculation was made by 
\citet{jeffery07}, that the iron group element enhancements
may be disrupted by the atmospheric motions as pulsations
build up to some level. They note that since $p$-modes mostly involve
vertical motion, while $g$-modes are dominated by horizontal motion,
it is possible that $p$-modes are more effective at redistributing the
iron group elements out of the driving zone. This could explain the
observation that most cool EHB stars are $g$-mode pulsators
while at the hotter end of the branch most are not \citep{green03}.
It may also be that the shortage of pulsators is due to observational
biases, either that the pulsations have too low amplitudes, or that they
have too high $\ell$ to be easily detectable photometrically, an issue
that needs further investigation.
To this end it also is important to publish observational constraints on
stars that were not found to be variable in order to guide future more
sensitive surveys.

Here we present the final results of a long-term programme to detect
new short-period sdBVs using the Nordic Optical Telescope ({\sc not}) on La Palma.
% In the section below we will describe the target selection procedure, and give
% a chronology of the survey with a summary of results published to date.
% In section 3 we describe the observational method, and in section 4 we
% summarise the results for the stars in our sample that were not observed
% to show significant variability, indicating the detection limit in each case,
% and providing a discussion of some borderline cases.
% In section 5 we report our results on four V361\,Hya stars not presented
% before, bringing the total number of short period sdBV stars discovered
% with the {\sc not} up to 20.
% This brings the total number of sdB stars known to display short period 
% pulsations up to 49, as summarised in section 6.

\begin{table}[t]
\caption[]{Observation runs at the Nordic Optical Telescope contributing to
this survey.} \label{tbl:obstime}
\begin{tabular}{llcrl} \hline\hline
ID  & Dates & $N_n$ & $N_t$ & Observers \\ \hline\\[-2.5mm]
N1 & 1999 July 19--23 & 4.0 & 13 & JES, RS, JMGP \\
N2 & 1999 Oct.~14--19 & 5.0 & 31 & JES, RH\O \\
N3 & 2000 July 7--11  & 4.0 & 25 & RH\O, RS \\
N4 & 2000 Oct.~4--8   & 4.0 & 38 & JES, RH\O \\
N5 & 2001 July 25--29 & 4.0 & 22 & RH\O , RS \\
N6 & 2002 Oct.~10--15 & 0.5 &  4 & JES$^\mathrm{a}$ \\
N7 & 2004 June 4 -- 8 & 4.0 & 44 & JES, RH\O \\
No & 2005 Feb.~15--20 & 5.0 &  0 & JES$^\mathrm{b}$ \\
N8 & 2006 Dec.~10--15 & 0.3 &  4 & JES$^\mathrm{a}$ \\
N9 & 2007 Dec.~14--16 & 3.0 & 14 & RH\O, JHT$^\mathrm{c}$ \\
X1 & 2004 Oct.~15--17 & 0.3 &  4 & JES, RH\O$^\mathrm{d}$ \\
S1 & 2004 May 21--24  & 4.0 & 20 & RO \\ 
S2 & 2005 May 30--June 1 & 4.0 & 15 & RO \\
S3 & 2007 Oct.~20--24 & 3.0 & 11 & RO \\
S4 & 2008 Sep.~22-25  & 3.0 & 20 & RO \\
S5 & 2008 July 26--30 & 2.0 & 67 & RH\O$^\mathrm{e}$ \\
S6 & 2009 Jan.~24--27 & 3.0 &  5 & RO \\
T1 & 2001 Feb.~24    & 0.6 &  7 & RH\O$^\mathrm{f}$ \\
T2 & 2001 April 10, 13   & 2.0 & 24 & RH\O$^\mathrm{e}$ \\
T3 & 2002 Sep.~13--14& 0.2 & 4 & RH\O, JHT$^\mathrm{f}$ \\
T4 & 2003 Feb.~10--19& 0.5 &  5 & RH\O, JHT$^\mathrm{f}$ \\
T5 & 2004 Jan.~13    & 0.5 &  9 & RH\O, JHT$^\mathrm{f}$ \\
T6 & 2005 May 5      & 0.7 & 16 & RH\O$^\mathrm{f}$ \\
T7 & 2006 May 9      & 0.7 & 15 & RH\O$^\mathrm{f}$ \\
T8 & 2008 Feb 24, 26 & 0.2 &  2 & RH\O$^\mathrm{f}$ \\
ST & 2006 June 10    & 0.5 &  2 & JMGP$^\mathrm{g}$ \\[0.5mm] \hline\\[-2.5mm]
\multicolumn{2}{l}{Total} & 59.0 & 421 & \\ \hline
\end{tabular}\\[2mm]
The ID column gives our run identifier, used to refer to a particular
observation run in Table~\ref{tbl:limits}
and when discussing particular observations.
$N_n$ gives the number of nights used for this survey programme.
For nights allocated to the survey, the number includes all 
nights, regardless of whether any observations were actually made or not.
For nights where this programme served as a backup for another,
only the fraction of nights actually used for this programme is listed.
The $N_t$ column shows the number of targets observed. Note that reobservations
and observations of sdO stars (to be analysed in a separate paper) are included
in these totals. The notes indicate\\
$^{\mathrm{a}}$ Run primarily dedicated to an unrelated target.\\
$^{\mathrm{b}}$ Run lost due to snow and ice.\\
$^{\mathrm{c}}$ Run primarily for follow-up of our pulsators.\\
$^{\mathrm{d}}$ {\sc Xcov}\,24 multi-site campaign.\\
$^{\mathrm{e}}$ Run primarily dedicated to search for sdO stars.\\
$^{\mathrm{f}}$ Run primarily for technical investigations.\\
$^{\mathrm{g}}$ Spanish service time.
\end{table}

\section{The survey}

Shortly after the discovery of the first pulsating sdBs by the South African
group \citep{kilkenny97,koen97,stobie97,odonoghue97}, it was decided to
initiate a programme to search for such pulsators using the 2.5\,m
{\sc not} on La Palma.
The principal aim was to extend the number of sdB pulsators,
determine the fraction of variables within this class of stars,
and explore the boundaries of their instability region.
The South African group had at the time discovered and confirmed pulsations
in 14 objects from a sample of around 600 spectroscopically confirmed
hot subdwarf stars, and a Canadian team found four pulsators in 74 stars
after selecting candidates based on \teff\ \citep{billeres02}.
Since those samples were overlapping considerably, two of the pulsators
were found in both samples.
To avoid searching an already depleted sample, we drew our
targets primarily from a new spectroscopic study targeting sdB star
candidates from the Hamburg-Schmidt (HS) survey \citep{hagen95}, although
it was complemented with targets from other surveys as explained below.

The complete list of observational runs fully or partly used for this survey
is presented in Table~\ref{tbl:obstime}.
In Table~\ref{tbl:limits} we provide coordinates and observational limits
on all sdB stars checked for variability during the survey, except those
that were found to pulsate or vary due to eclipses and reflection effects.
The eclipsing and reflection variables are published in separate papers,
and the 20 pulsators are listed in
Table~\ref{tbl:sdbvars}, together with pulsators found by other surveys.

\subsection{Chronology and results published to date}

The rapid pulsations in sdBVs are a challenge to observe with standard CCD
instrumentation
due to the pulsation periods being comparable to the typical readout times of
a full CCD frame. An essential part of this project was the design of a system
capable of efficient windowed CCD photometry \citep{ostensen00}.
A dedicated camera was built for this purpose, the Troms\o\ CCD Photometer
(TCP), and the control system implemented for this camera was adopted to both
the \hirac\ and \alfosc\ cameras at the {\sc not}.

The first search run was made in the summer of 1999
(c.f.~Table~\ref{tbl:obstime}), using the Troms{\o}-Texas
Photometer (TTP) and revealing the first sdB pulsator from this programme
\citep[Paper {\sc i}]{silvotti00}.
Already the same autumn the CCD system was operational, and the improved
detection efficiency allowed us to reach acceptable noise levels in a much
shorter time, thereby resulting in the discovery of three new pulsators in
only five nights of observations \citep[Paper {\sc ii}]{ostensen01a}.
Two observation runs in 2000 led to the discovery of
four more pulsators \citep[Paper {\sc iii}]{ostensen01b}.
At this time we started to run out of targets from the HS survey at some
RA's, and supplemented our list with stars from the PG survey \citep{pgcat}.
This was a concern, as many of the known sdB stars had already been targeted
by the South African group, and we expected our detection efficiency to drop
since we avoided targeting stars already reported to pulsate.
Still, the sample was not completely depleted, and during the summer 2001
run we discovered pulsations in \object{QQ\,Vir} (\object{PG\,1325+101})
and \object{EP\,Psc} which originated from the HS sample
(\object{HS\,2303+0152}) but also appear in the PG catalogue as
\object{PG\,2303+019}  \citep[Paper {\sc iv}]{silvotti02a}.

To get a sample that was both sufficiently large to properly determine
the extent of the sdB instability region and the fraction of pulsators
across this region, and at the same time undepleted, it was decided to
supplement our target lists with stars from the SDSS survey \citep{SDSS}.
The first three SDSS stars were observed in October 2002, and one of them,
\object{SDSS\,J171722.10+580559.9} or
\object{J1717+5805} for short,
was found to pulsate \citep[][Paper {\sc v}]{solheim04}.

In spite of the initial success of our programme, the
Nordic time-allocation committee started turning down our proposals
for further search time, and the last regular search run on the
original programme was allocated for February 2005, but was completely
lost due to bad weather. The last successful search run was made in June
2004, resulting in three new pulsators.
These are \object{PG\,1419+081},
\object{SDSS\,J144514.93+000249.0} (\object{J1445+0002}) and
\object{SDSS\,J164214.21+425234.0} (\object{J1642+4552}).
Only brief sequences were made on these objects, and the
results were presented in \citet[][Paper {\sc vi}]{solheim06}.

At this time a collaborative search was initiated with a Spanish team
(RO, AU, FPH, CRL), which permitted additional access to the {\sc not} through
Spanish time. With the Spanish group, we were able to explore a
larger number of stars drawn from the Subdwarf Database
\citep{ostensen04,ostensen06}. The first pulsator discovered with the
{\sc not} on Spanish time was another star drawn from the SDSS sample
\object{PG\,1657+416} (\object{SDSS\,J165841.83+413115.6}), and
was presented by \citet[Paper {\sc vii}]{oreiro07}.
Observation runs allocated on Nordic time are indicated by run numbers
prefixed by N, and on Spanish time by S in Table~\ref{tbl:obstime}.

Nordic time still continued to be granted for follow-up programmes on various
pulsating stars, including one of the most
interesting pulsators discovered during this programme, \object{V391\,Peg},
which was demonstrated to be host to a planet by \citet{silvotti07}, by using
the orbital period modulation introduced on the two main pulsation modes.
%to determine the minimum mass and orbital period of the planet.
A few unexplored pulsator candidates were observed during these runs
(N6, N8, N9 in Table~\ref{tbl:obstime}),
and also during the {\sc wet Xcov\,24} campaign on the pulsating sdB star
UY\,Sex \citep[][run X1]{maja06},
whenever the prime target was not reachable. 
Upgrades and tests on the fast CCD photometry system also gave several
opportunities to observe some additional targets during technical time
at the {\sc not} (listed as T1 to T8 in Table~\ref{tbl:obstime}).
One new pulsator, \object{PG\,1033+201}, was found during run T7.

The limitations on telescope time from 2004 onward forced us to abandon the
faint SDSS targets. Instead, we focused on detecting more pulsators in
the cool region where the particularly interesting objects \object{V391\,Peg}
and \bal\ had been found.
The discovery of hybrid pulsations in \object{DW\,Lyn} by \citet{schuh06}
provided further encouragement for increasing the sample in this region
of the instability strip.
Although no more such cool pulsators were detected,
a total of five more pulsators in the hotter end of the instability strip
were found.  One, from the BG sample \citep[Bok-Green,][]{green08},
was recently published in \citet[Paper {\sc viii}]{oreiro09}, and
four are presented for the first time in this paper. 

In addition to these 20 pulsators our photometric observations
have revealed several of the relatively rare sdB stars with M-dwarf
companions from their strong reflection effect.
The discovery of an eclipsing sdB+dM system, \object{HS\,0705+6700},
first detected during run N4
was presented by \citet{drechsel01}, and the non-eclipsing sdB+dM
system \object{HS\,2333+3927} discovered during run N2 was presented
by \citet{heber04}. Two more sdBs with dM companions were found during
run S2: the eclipsing \object{HS\,2231+2441} \citep{ostensen07}, and
\object{HS\,2043+0615} \citep{hs2043}.

Our survey was not originally designed to detect long period variability,
since we aimed for short photometric sequences with high S/N.
However, the large photometric
amplitudes of sdB+dM binaries produce a strong trend in the
differential photometry, of which we saw several examples.
However, since follow-up of such long-period variable objects can easily
be done with smaller telescopes using long integration times, no follow-up
was attempted on these stars with the {\sc not}.
The long period $g$-mode sdBVs also pulsate on time-scales around one hour,
but with much lower amplitudes.
As they had yet to be discovered at the start of our survey,
our observation strategy was not designed with these objects in mind.

\begin{table}[t]
\caption{Analysis for eight stars from the HS survey, not included in
\citet{edelmann03}. }
\label{tbl:hsnew}
\begin{center}
\begin{tabular}{lclccc} \hline\hline
Name & \mpg & Class & \teff & \logg & \logy \\
     & mag  &       & kK    & dex   & dex   \\\hline\\[-2.5mm]
\multicolumn{6}{l}{Observed with CA/\twin} \\
\object{HS\,1733+4540}         & 15.1 & sdOB & 37.5 & 5.6 & $-$3.0 \\
\object{HS\,1909+7004}         & 15.4 & sdB  & 36.5 & 5.4 & $-$2.9 \\[0.5mm]
\multicolumn{6}{l}{Spectra courtesy of \citet{YAS_PhD}} \\
\object{HS\,0026+0439}$^1$     & 14.8 & sdB  & 36.7 & 5.8 & $-$3.0 \\
\object{HS\,0042+0927}$^1$     & 16.0 & sdB  & 30.2 & 5.6 & $-$2.9 \\
\object{HS\,2254+0640}$^1$     & 15.2 & sdB  & 29.1 & 5.6 & $-$2.7 \\
\object{HS\,2320+0840}$^{1,2}$ & 14.6 & sdB+ & 30.0 & 5.8 & $-$2.6 \\
\object{HS\,2323+0459}$^1$     & 14.8 & sdB  & 29.3 & 5.4 & $-$2.8 \\
\object{HS\,2334+0144}         & 15.5 & sdB  & 30.8 & 5.6 & $-$2.6 \\
% HS\,2249+0026 & 16.1 & sdOB & 36.4 & 5.50 & ? \\ % no spectrum!
% HS\,0127+3146 & 14.4 & sdB+ & 33.5 & 5.2 & -3.0 \\ % why here?
\hline\\[-6.0mm]
\end{tabular}\end{center}

\mpg\ gives the photographic magnitude from the HS survey plates,
`Class' our spectroscopic classification, followed by the
effective temperature (\teff), surface gravity (\logg), and
atmospheric helium abundance ($\log y$\,=\,\lheh), from our model fits.
The sdB+ classification indicates a composite spectrum. The notes on
the names imply:\\
$^{1}$ Also found in the PG survey \citep{pgcat}.\\
$^{2}$ Also found in the KUV survey \citep{kuv}.
\end{table}

\subsection{The HS/HE/SPY samples}\label{sect:hssamp}

When our programme was initiated, the main source of targets was the HS survey,
but we also included targets from
the equatorial part of the Hamburg-ESO (HE) survey \citep{wisotzki96}.
Follow-up spectroscopy made at Calar Alto with the \twin\ spectrograph
\citep{heber99,edelmann03} allowed us to preselect candidates with
effective temperatures and gravities in the domain predicted for the 
pulsational instability \citep{charpi97}.
This provided us with a sample of pulsator candidates that was
both tuned to the instability region and relatively undepleted
by previous surveys. The latter was particularly
important since the South African group had already explored
a considerable fraction of the sdB stars known in the literature
\citep[and compiled in the {\em Catalogue of spectroscopically identified
hot subdwarf stars},][]{kilkenny88}.
However, it has turned out that the HE survey has a significant overlap
with unpublished parts of the Edinburgh-Cape (EC) blue object survey
\citep{ECsurvey}, which was one of the main sources surveyed by the
South African group. This can explain
why out of 80 stars from the HS survey and 20 stars from the HE survey
observed during the initial five runs \citepalias[as reported in][]{solheim04},
pulsations were found in eight HS stars and in none of the HE stars.
Due to the difficulty of establishing accurate effective
temperatures and gravities for sdB+F--K stars, these were
not included in our initial sample, and were only added later when the
priority sample started to become depleted.
Thus, unlike the South African group, who found that most of their
first pulsators were in binary systems, due to the selection biases involved
only one of our initial 10 were. Note that another composite,
our first pulsator \object{LM\,Dra},
is in a well-separated visual binary with an F3 companion, and
therefore not considered to be an sdB+F system since the stars cannot have
interacted during their evolution.

Only eight stars from the HS sample observed during this survey
have not yet been published elsewhere. Their coordinates are listed in
the on-line table together with those of all the other targets
(Table~\ref{tbl:limits}),
and their physical parameters are listed in Table~\ref{tbl:hsnew}.
We computed the stellar atmospheric parameters (effective temperature,
surface gravity, and photospheric helium abundance) by fitting model
atmosphere grids to the hydrogen and helium lines visible in the
spectra. The procedure used for this fitting was the same
as that of \citet{edelmann03}, but using only the LTE models described in
\citet{heber00}, as the purpose of our modelling was only to
establish which stars are roughly within the sdB instability region
rather than to establish precise physical parameters.

The parameters of \object{HS\,1909+7004} are uncertain since the spectrum is
of low resolution and the fit is quite poor, but the fits to the other stars
appear to be quite good.
\object{HS\,2320+0840} has a very strong companion that contributes
clearly visible line features in the spectrum.
For this reason, we performed a spectral decomposition,
subtracting a main sequence model from \citet{munari05}
with \teff\,=\,6.5\,kK, \logg\,=\,4 and solar metallicity, which
we found to contribute about 40\%\ of the flux at 5\,000\,\AA.

A total of 46 objects from the HE survey were observed.
For 23 of these the physical parameters were determined by \citet{EdelmannPhD},
and none of these were found to pulsate.
Cross-referencing them with the literature reveals that only nine
are new, the rest are subdwarfs with identifications from
the Montreal-Cambridge-Tololo (MCT) Survey \citep{MCTCat}, or earlier
surveys compiled by \citet{kilkenny88}.

A number of white dwarf candidates from the HE survey were 
observed as part of the {\sc spy} survey \citep{spy}
and published by \citet{lisker05}, 
where physical parameters are provided also for the composite objects.
23 HE stars from the {\sc spy} sample were observed, and two were found to
pulsate. Also, four HS stars from this sample were observed, and one
more pulsator was found.
These three new variables are presented in Sect.~\ref{sect:pulsators}.
\citet{lisker05} also present a list of sdB stars formerly misclassified
as white dwarfs. We observed four of these, but did not detect signs of
variability.\footnote{Actually, one more of the fourteen stars in this list of
misclassified WDs is included in our sample: \object{Ton\,S\,155}
appears in \citet{EdelmannPhD} as \object{HE\,0021--2326}.}

\begin{table}[tb]
\caption{The stars from the SDSS survey observed during this programme,
excluding the pulsators listed in Table~\ref{tbl:decomp}. }
\label{tbl:sdss}
\begin{center}
\begin{tabular}{lcllll} \hline\hline\\[-2.5mm]
Name                   & $g'$ & Class & \teff & \logg & \logy\\
                       & mag  &       & kK    & dex   & dex  \\\hline\\[-2.5mm]
PB\,5916$^1$           & 15.3 & sdB  & 30.1 & 5.7 & --2.3\\ %check
PG\,0812+482$^2$       & 15.0 & sdB  & 25.0 & 5.4 & --3.0\\ %chk
PG\,0826+480$^2$       & 15.8 & sdB  & 26.1 & 5.5 & --3.1\\ %chk
PG\,1100--008$^2$      & 16.3 & sdB+ & 29.2 & 5.5 & --2.6$^\dag$\\ %chk
PG\,1136--003$^{1,2}$  & 14.2 & sdB  & 31.6 & 5.6 & --2.7\\ %ok
PG\,1249+028$^{1,2,3}$ & 15.4 & sdB  & 30.6 & 5.6 & --3.0\\ %chk
J12596--0039$^4$       & 16.8 & sdOB+& 34.4 & 5.6 & --2.2$^\dag$\\ %chk
PG\,1315+013$^{1,2}$   & 16.6 & sdB  & 26.6 & 5.3 & --2.3\\ %chk
J13320+6733$^4$        & 17.0 & sdOB & 36.1 & 6.0 & --1.1\\ %chk
J13516+0234$^1$        & 17.1 & sdB+ & 34.7 & 5.3 & --3.0$^\dag$\\ %chk
PG\,1403+019$^{1,2}$   & 15.6 & sdB  & 27.7 & 5.5 & --1.9\\ %chk
J14086+6531$^1$        & 17.4 & sdB  & 30.4 & 5.6 & --3.0\\ %chk
J14236+0149$^1$        & 17.1 & sdB  & 29.5 & 5.7 & --2.3\\ %chk
PG\,1422+035$^2$       & 16.2 & sdOB & 35.1 & 5.8 & --1.5\\ %chk
J15107+0409$^4$        & 16.9 & sdOB & 34.7 & 5.7 & --1.6\\ %chk
J15131+0114$^1$        & 17.1 & sdB  & 28.0 & 5.6 & --3.0\\ %chk
EGGR\,491$^{1,5}$      & 16.8 & sdB  & 30.6 & 5.6 & --3.0\\ %chk
SBSS\,1544+568$^1$     & 16.8 & sdB  & 27.2 & 5.2 & --2.0\\ %chk
J15481--0049$^1$       & 16.4 & sdB+ & 32.7 & 5.6 & --3.0$^\dag$\\ %chk
J15513+0649            & 15.9 & sdOB+& 34.3 & 5.0 & --2.1$^\dag$\\
J15564+0113$^1$        & 16.0 & sdB  & 30.7 & 5.7 & --3.1\\
J16026--0012$^1$       & 17.5 & sdB  & 34.0 & 5.8 & --2.1\\
J16165--0038$^1$       & 16.7 & sdOB & 36.0 & 5.7 & --1.8\\
J16331+0032$^1$        & 16.9 & sdB  & 31.7 & 5.8 & --3.3\\
J16347--0053$^1$       & 16.9 & sdB  & 29.1 & 5.5 & --2.7\\
J16420+4403$^1$        & 16.7 & sdB  & 28.8 & 5.2 & --2.8\\
J16443+4523$^1$        & 17.1 & sdB  & 33.1 & 5.8 & --2.0\\
PG\,1653+633$^{1,2}$   & 15.9 & sdOB & 36.1 & 5.9 & --1.5\\
J17144+6147$^1$        & 16.6 & sdB+ & 33.4 & 5.6 & --3.4$^\dag$\\
SBSS\,1715+556$^{1,6}$ & 16.9 & sdB  & 30.3 & 5.4 & --3.2\\
J17165+5751$^4$        & 18.0 & sdOB & 34.9 & 5.8 & --0.8\\
J20440--0511$^1$       & 17.3 & sdB  & 29.9 & 5.4 & --3.2\\
J20457--0543$^1$       & 17.8 & sdB  & 34.9 & 5.4 & --1.7\\
J20573--0626$^1$       & 16.5 & sdB  & 28.0 & 5.1 & --2.5\\
J21531--0719$^1$       & 16.9 & sdB  & 32.0 & 5.9 & --2.0\\ % PHL189
\hline\\[-6.0mm]
\end{tabular}\end{center}
The columns are as in Table~\ref{tbl:hsnew}, except that Sloan $g'$ magnitudes
are provided provided as indicator of the signal level.
A plus sign after the class marks the presence of a cool companion from
features in the spectrum, the slope of the spectrum or $z$-band
excess. The notes indicate:\\
$^{1}$ Correctly identified as sdB stars in \citet{eisenstein06}.\\
$^{2}$ Classified as hot subdwarf stars in \citet{pgcat}.\\
$^{3}$ Object misnamed in the PG catalog; should be 
       \object{PG\,1249--027}.\\
$^{4}$ Listed as sdO stars in \citet{eisenstein06}.\\
$^{5}$ Misclassified as a white dwarf in \citet{wdcat}.\\
$^{6}$ Identified as an sdB star in \citet{stephanian01}.\\
$^{\dag}$ Parameters for composites are less certain than for singles.
\end{table}

\begin{table*}[t]
\centering
\caption{Spectral decomposition of the binary pulsators from the SDSS sample.}
\label{tbl:decomp}
\begin{tabular}{lcccclcccccccc} \hline\hline
     & \multicolumn{4}{c}{Companion signature} &       & \multicolumn{4}{c}{Primary} & \multicolumn{4}{c}{Secondary} \\
Name                  & \ion{Ca}{ii} & CH & \ion{Mg}{i} & \ion{Na}{i} & $E(B-V)$& $f_{5000}$ & \teff & \logg & \logy &
                               $f_{5000}$ & \teff & \logg & [M/H] \\ 
                      & H+K & G &  b & D & mag & & kK & dex & & & kK & dex & \\ \hline\\[-2.5mm]
\object{J1717+5805} & \chk &      & \chk & \chk & 0.025 & 0.90 & 34.4 & 5.75 & $-$1.8 & 0.10 & 5.0 & 4.0 & +0.0 \\
\object{PG\,1657+416}   & \chk & \chk & \chk & \chk & 0.021 & 0.74 & 32.2 & 5.80 & $-$2.0 & 0.25 & 5.5 & 4.0 & +0.0 \\
\object{PG\,1419+081}   & \chk & \chk & \chk & \chk & 0.028 & 0.86 & 33.3 & 5.85 & $-$1.8 & 0.14 & 5.5 & 4.0 & +0.0 \\
\object{J1445+0002}     & \chk &      & \chk & \chk & 0.11  & 0.83 & 35.9 & 5.75 & $-$1.6 & 0.17 & 6.0 & 4.0 & +0.0 \\
\object{J1642+4552}     & \chk & \chk & \chk & \chk & 0.011 & 0.53 & 32.4 & 5.80 & $-$2.0 & 0.47 & 6.0 & 4.0 & $-$1.5 \\
\hline
\end{tabular}
\end{table*}

\subsection{The SDSS sample}\label{sect:sdsssamp}

More than 16\,000 spectra of UV excess stars were downloaded from the
SDSS survey \citep{SDSS} up to data release 6 (DR6), and the hot subdwarfs
were identified by inspection.
The intermediate resolution ($\sim$1\,\AA) SDSS spectra cover all
wavelengths from 3\,800 to 9\,000\,\AA, ideal for classification purposes.
We identified about 900 objects as hot subdwarf stars and classified them
into sdB, sdO, sdOB, He-sdB, He-sdO, and He-sdOB, 
depending on whether hydrogen or helium dominates the spectrum, and if
they contain lines from \ion{He}{i}, \ion{He}{ii} or both.
318 of the sdB stars
were classified as single sdB stars and 165 as sdB+F--K binaries (34\%) based on
spectroscopic signatures and their position in ($u'-g'$,\,$g'-r'$) colour-colour
space.  A similar fraction was found for the sdOB stars;
78 single and 35 binaries (31\%).  All these stars were
fitted to model spectra to determine their position in the \teff/\logg\
plane, using the Balmer lines \ion{H}{$\alpha$} to \ion{H}{$\eta$}, as well as
the most prominent helium lines at 4472\,\AA\ and 4026\,\AA\ for the sdB stars,
and including \ion{He}{ii} 4686 for the sdOBs.
The formal fitting errors are about 250\,K for \teff, 0.05\,dex for
\logg, and 0.1\,dex for \logy\,=\,\lheh\ at a $g'$ magnitude around 16.0,
but increase by a factor of two
for the fainter stars between $g'$\,=\,17.5 and 18. For the stars with
F--G companions the uncertainties are larger, but not so large
that the parameters are not useful for establishing whether or not the stars
are located within the instability region.
We note also that for the hottest stars in the sample, NLTE effects start to
become significant \citep{napiwotzki97}, and that the LTE models can
underestimate the temperatures by up to 1000\,K in some cases.

Only 40 stars from this large sample were observed due to observing time
restrictions. Five were found to pulsate and are listed in
Table~\ref{tbl:decomp} with additional details in Table~\ref{tbl:sdbvars}.
The remaining 35 objects are listed in Table~\ref{tbl:sdss}, with the physical
parameters we derived from the SDSS archive spectra.
The sample includes eleven spectroscopic binaries, and curiously all five pulsators
found from the SDSS sample were among these. The physical parameters for the
binaries have considerable systematic shifts depending on the contribution from the main
sequence companion. For this reason we have made a spectral decomposition of
the five pulsators in order to place them reliably in the \teff/\logg\ plane.
The details of the spectroscopic decomposition are given in
Table~\ref{tbl:decomp}.
The six non-pulsating sdB+F--K systems were fitted as if they were single stars.

It is non-trivial to estimate the spectroscopic parameters of the
companion, but a reasonable compromise can usually be made that approximates
the spectral contributions from the main sequence companion to the
\ion{Ca}{ii} H and K lines as well as the g-band and the \ion{Mg}{i} lines.
For the spectral decomposition we used the SDSS spectra processed with the
DR7 pipeline, which have a significantly improved flux calibration compared
with earlier data releases. After dereddening the spectrum by the
$E(B-V)$ reddening coefficients provided by the \citet{schlegel98} dust maps,
the flux was
fitted with one of our subdwarf model spectra and a main sequence model
spectrum from \citet{munari05}. The parameters of the subdwarf are then refitted
to the observed spectrum after subtraction of the main sequence model, and the
procedure was iterated until it converged.
Note that the reddening coefficients are for lines of sight to infinity and
that we have no guarantee that the particular subdwarf under consideration is
not in front of the bulk of the dust. There is also a significant uncertainty
in the $E(B-V)$ values due to the limited resolution of the \citet{schlegel98}
maps and the lumpiness of the interstellar medium.
But due to the low resolution and limited signal in the SDSS spectra, it
is hard to make a complete disentanglement of the spectra while leaving
the reddening as a free parameter, so we will stick with the
dust map values.
In the case of \object{J1445+0002}, which has the highest
reddening coefficient of the five, it is quite clear that the \ion{Ca}{ii} lines
are too narrow to originate from the main sequence companion, and no solution
can be found that removes the \ion{Ca}{ii} contribution from the
subdwarf spectrum. Therefore, the \ion{Ca}{ii} signature is mostly
interstellar in origin, consistent with the high reddening value.
The strongest companion by far is that of \object{J1642+4552}, in which it
contributes almost half the light at $\lambda$\,=\,5000\,\AA\ ($f_{5000}$ in
Table~\ref{tbl:decomp}).  The system is also in
a region of very low extinction, reducing the ambiguity of the spectral
decomposition.  The strong companion and good signal in this particular
SDSS spectrum allow us
to make some further constraints on the companion. It is clear that a
solar metallicity makes a far stronger contribution to the metal
bands than allowed by the observed spectrum. A better fit is achieved
with a spectrum metal depleted relatively to solar by [M/H] = $-$1.5.

The temperatures derived here are somewhat lower than the temperatures
we originally used to select the stars, 
but not enough to make a difference with respect to the sample selection,
as we suspected \citepalias{oreiro07}.
For J1445+0002 the temperature drops from
37.6 to a more reasonable 35.9\,K. It is suspicious that the derived
\logg\ values for all the five composite stars are so similar
(within $\pm$\,0.1 dex), and might be due to degeneracies when
fitting so many free parameters to spectra with rather low resolution.
It has also been noted that more reliable estimates of the surface
gravity are obtained when the spectra cover all the high order Balmer
lines, a requirement the SDSS spectra do not satisfy.

\subsection{The Bok-Green sample}

Eleven stars in our survey were selected from a spectroscopic campaign
to obtain a large unbiased sample of hot stars based on {\sc 2mass}
colours, undertaken by \citet{green08} with the University of Arizona 2.3\,m
Bok telescope.
Physical parameters were derived from a subsample of 89 stars
from this campaign by \citet[][Appendix C]{WinterPhD}. 
None of the stars we observed have reliable spectroscopic classifications in the
literature, but three have been listed as faint blue stars by earlier
surveys (\object{Ton\,930}, \object{FBS\,1133+754}, and \object{FBS\,1224+780}).
However, none of these were known to be hot subdwarf stars, so we consider all
eleven as belonging to an undepleted sample.
One of the sdBs in the BG sample was found to pulsate,
\object{2M0415+0154} = \object{{\sc 2mass} J04155016+0154209}, as
reported in \citetalias{oreiro09}.
The remaining seven stars are listed with 2M designation in Table~\ref{tbl:limits}.

\subsection{Literature}

To supplement our sample with targets to fill all RAs we added stars with
temperatures and gravities published in the literature. 
Most of these additional objects are well-known subdwarfs from the PG
survey and are included in many studies.
The largest sample used was that of \citet{saffer94}, where 24 targets were
surveyed and one found to pulsate (\object{LM\,Dra}).

Seventeen PG stars from the radial velocity survey of \citet{maxted01},
and four stars from the HST study of \citet{heber02} were
also included in our sample, but none were found to be variable.

Rough temperature estimates can also be made on the basis of photometric
measurements alone, and several works provide such estimates.
The survey of \citet{beers92} includes temperatures computed from UBV
colours, and we observed eight of these stars.
A handful of composite sdB+MS stars from the list of \citet{allard94},
where their estimate of the temperature for the primary lie within the
instability region, were also observed.
Another four stars were targeted based on the temperature estimates
of \citet{bixler91}.
Similarly, we observed stars from the sample of \citet{moehler90}, after
estimating the temperatures based on their Str\"omgren photometry
and the spectral features visible in their spectroscopic atlas.
In total, about fifty stars in the sample were observed based on such
photometric temperature estimates, and only one, \object{PG\,1419+081},
was found to vary.  The SDSS spectroscopy of this object was released
after our discovery, allowing us to refine the original colour temperature
estimate and correctly identify the star as an sdB+G5 composite.
We found that the colour temperature \citep[\teff\,=\,33.8\,kK;][]{beers92}
is encouragingly close to our model fit value (\teff\,=\,33.3\,kK).

If we look at the complete sample from \citet{beers92} and consider only
the stars classified as sdB with temperatures above 27.5\,kK we are left
with 23 stars, of which we have observed nine and found one variable. But
\object{EC\,20338--1925} is also in this sample and was in our original list,
although excluded after pulsations were discovered by the South African team
\citep[published by][but preliminary results from the 1998 discovery
were distributed much earlier]{kilkenny06}.
Another star in the sample is \object{BPS\,CS\,22169--1},
analysed by \citet{geier09} and found to have \teff\,=\,39.1\,kK,
much hotter than the 33.9\,kK estimate by \citet{beers92},
and hotter than the predicted sdBV instability region.
The star \object{BPS\,CS\,22890--94} = \object{PG\,1525+024} was not
observed since the colour temperature estimate by \citet{beers92}
placed it below our cut at \teff\,=\,22.9\,kK. However, recent SDSS
spectroscopy gives us \teff\,=\,27.9\,kK, which would have placed the
object inside our cuts, further emphasising the uncertainties
associated with using colours as a temperature estimator.

\begin{table*}[t]
\caption{Truncated version of the catalogue of sdB stars observed during our search programme.
}
\label{tbl:limits}
\begin{center} \small
\begin{tabular}{llccccrrrrl} \hline\hline
Target Name & Other Name & RA & Dec & \mpg & Run &$N_p$&$\Delta t$&$\sigma$& $A_\mathrm{max}$ & Var \\
            &            & J2000 & J2000 & mag   &     &     &      s   & mma & mma & \\ \hline\\[-2.5mm]
SB\,7           & MCT\,0000--1637  & 00:03:24.37 & $-$16:21:06.3 & 12.7 & S5 &  60 & 20 & 0.60 & 1.26 & NOV \\
HE\,0002--2648  & MCT\,0002--2648  & 00:05:09.58 & $-$26:31:48.4 & 15.6 & S5 &  39 & 20 & 1.01 & 2.07 & NOV \\
HE\,0004--2737  & MCT\,0004--2737  & 00:06:46.26 & $-$27:20:53.4 & 13.7 & N4 &  60 & 20 & 0.44 & 0.95 & NOV \\
HE\,0007--2212  & MCT\,0007--2737  & 00:09:45.91 & $-$21:56:14.4 & 14.9 & S5 &  43 & 10 & 0.79 & 1.48 & NOV \\
MCT\,0008--1245 & ...              & 00:11:25.04 & $-$12:28:45.9 & 14.3 & S4 &  93 & 15 & 0.59 & 1.04 & NOV \\\hline\\[-2.5mm]
KUV\,04233+1502 & ...              & 04:26:08.00 &   +15:08:29.01 & 14.3 & S4 & 129 & 20 & 0.33 & 1.01 & \\
KPD\,0716+0258  & ...              & 07:18:57.85 &   +02:53:14.67 & 14.7 & T4 & 270 & 30 & 0.25 & 0.78 & \\
KUV\,07301+3659 & ...              & 07:33:25.62 &   +36:52:57.83 & 14.8 & S4 & 134 & 18 & 0.30 & 0.93 & \\
PG\,1248+164    & ...              & 12:50:50.23 &   +16:10:03.40 & 14.0 & T7 &  65 & 20 & 5.07 &16.61 & \\
PG\,1313+132    & ...              & 13:15:58.16 &   +12:57:40.42 & 14.1 & S6 &  96 & 20 & 0.95 & 3.35 & \\
J1351+0234      & ...              & 13:51:40.69 &   +02:34:29.21 & 17.1 & N7 &  70 & 30 & 0.91 & 2.93 & \\
PG\,1722+286    & ...              & 17:24:11.97 &   +28:35:26.90 & 12.8 & N1 &1191 &  5 & 1.01 & 3.08 & \\
PG\,1725+252    & ...              & 17:27:57.39 &   +25:08:35.70 & 12.7 & T2 &  61 & 10 & 3.38 &12.39 & \\
HS\,2029+0301   & ...              & 20:31:52.63 &   +03:11:17.39 & 15.6 & N1 & 952 &  5 & 2.13 & 6.41 & \\
KPD\,2215+5037  & ...              & 22:17:20.62 &   +50:52:58.09 & 13.6 & N5 &  45 & 25 & 0.69 & 2.39 & \\
\hline
\end{tabular}\end{center}
The full table with 285 entries is available in electronic form from CDS, via
anonymous ftp to cdsarc.u-strasbg.fr or via
http://cdsweb.u-strasbg.fr/cgi-bin/qcat?J/A+A/???/???.
Only the first five entries in the table are listed, plus the ten stars that
have peaks between 3\,$\sigma$ and the 3.7\,$\sigma$ cutoff required for the
NOV flag (discussed in Sect.~\ref{sect:pcand}).
The stars in the sample for which pulsations were detected are
listed in Table\,\ref{tbl:sdbvars}.
The columns include target names, coordinates, a magnitude estimate
(usually photographic), the run ID from Table~\ref{tbl:obstime}, the
number of data points in the light-curve ($N_p$) used to compute the FT,
the sampling interval ($\Delta t$), the mean amplitude in the FT ($\sigma$),
the amplitude of the highest peak in the FT ($A_\mathrm{max}$), and a
non-variability flag.
\end{table*}

\section{Observations}

During our observations we systematically aimed at identifying
as many clear pulsators as possible, rather than trying to push the limit
on the pulsation amplitudes as low as possible. Thus, most of our detected
variables have rather high pulsation amplitudes, and the sample may still
contain objects with undetected amplitudes below 2 mma. Due to the shortness
of our typical photometric sequences, there may also be multiperiodic pulsators
among our null-detections, unidentified simply because their periods happened
to be canceling each other out in a beat phase at the short time they were
observed. The stars in Table~\ref{tbl:limits}
should therefore not be considered non-variable, rather, we prefer the term
{\em not observed to vary} (NOV).

For the aims of this campaign we consider a star as NOV if there is no peak
in its Fourier Transform (FT) with an amplitude higher than 3.0 times the
mean amplitude level ($\sigma$).
However, we require an amplitude of 3.7\,$\sigma$
\citep[99\%\ confidence level]{kuschnig97}
in order to call the object a pulsator. Objects that fall between these two
tiers are considered pulsator candidates until further observations can confirm
the detection.
Most pulsator candidates were reobserved in subsequent runs, but 11 objects
remain in Table~\ref{tbl:limits} with this classification,
and will be discussed separately in Sect.~\ref{sect:pcand}.

Only the frequency region 1.5--10\,mHz was used to
compute $\sigma$ homogeneously.
%, despite the Nyquist frequency varies from 12.5 to 50\,mHz.
The low frequency range was excluded to avoid the higher mean amplitude
level caused by small trends in the light-curve due to extinction effects
or sky transparency variations. 
% As noted below, wide photometric filters
% were used for the fainter stars during this program, which may preclude a
% proper extinction correction, especially when the target has a considerably
% different colour than the comparison stars, as is most often the case with
% the extremely blue sdBs.  Note, however, that the observations were designed to
% target most of the targets at low air
% masses, and usually for too short time to face extinction problems.

\subsection{PMT photometry}

The first observation run (N1) was made with the Troms\o-Texas 3-channel
photoelectric photometer, equipped with Hamamatsu R647 photomultipliers.
The data were reduced with the standard {\tt quilt} software package
developed for the {\sc wet} \citep[Whole Earth Telescope;][]{nather90}.
The reductions include smoothing of the sky background, and
correction for extinction and linear trends.
A total of 13 stars were observed, and one pulsator detected.
However, due to problems with the guiding many of the sequences were
rather poor, giving a $\sigma$ of about 2 mma
in the FT. Several stars showed a behaviour that could
be interpreted as pulsations, but was actually due to periodic
motion in the apertures.  For this reason three stars were reobserved
during the N2 run, and only the sequence with the lowest $\sigma$ is listed
in Table~\ref{tbl:limits}.
% Three stars from this run have too high noise levels (above 1\,mma) to
% qualify for an NOV classification, but have not been reobserved.

\subsection{CCD photometry}

Further {\sc not} observations were done with
the High Reso\-lution Adaptive Camera (\hirac) for the second run
(N2 in Table~\ref{tbl:obstime}), and the Andaluc\'ia Faint Object
Spectrograph and Camera (\alfosc) for all the remaining runs.
Both \hirac\ and \alfosc\ were equipped with Loral, Lesser thinned,
2048$\times$2048 CCD chips, up until January 2004, when \alfosc\ was
upgraded with an E2V CCD 42-40 device.
The sky area available for locating reference stars was limited to
$\sim$\,3.7$\times$3.7 and $\sim$\,6.5$\times$6.5 arcminutes$^{2}$ respectively
for each of the two cameras. 
All CCD observations used the multi-windowed
fast CCD photometry \citep{ostensen00}. Efficient fast 
photometry is easily achieved with this mode, since only small regions
(windows) of the chip are read. 
% In particular: one window for the target,
% one other free of stars to estimate the sky background, and several
% for reference stars depending on availability in the particular field.

Observations were made with either a Bessell $B$-band filter
\citep[{\sc not} \#74,][]{bessell}, for the brighter targets, and
a much wider filter
({\sc not} \#92) for the fainter ones.\footnote{
Filter details with transmission curves for all {\sc not} filters
can be found on their web-pages; {\tt http://www.not.iac.es/}.}
This filter (hereafter referred to as the $W$, or wide filter)
has the same centre as that of the $V$-filter (5\,500\,\AA),
but is a full 2\,750\,\AA\ wide, effectively covering all bands from
$B$ to $R$ with more than 90\% transmission.
The cycle times were mostly set to 20\,s, except for some of the brightest
objects, which can have cycle times as short as 5\,s, and the faintest ones,
which may have times up to 40\,s. Between 2.5 and 6.5\,s of the cycle times
are readout and shutter overheads, the exact number depending on the
number of reference stars selected and the size of the readout windows.

The CCD data were reduced on-line with the Real Time Photometry (RTP) 
program developed by one of us as part of his Ph.D--project
\citep[see also Paper~\sc{ii}]{mythesis}.
This software was particularly useful for the purposes of this survey, as it
performs real-time reduction of the data during sequencing,
displays the light-curve of the target, and computes the FT.
Thus, the noise level and the presence of clear variability could be 
checked at the telescope, although careful data reduction was also done afterwards.
The processing includes bias level removal, flat fielding,
sky subtraction, extinction correction and aperture photometry
using apertures that track each star's geometrical centre.
%The resulting limits are reproduced in table \ref{tbl:limits}.

The optimal aperture for each sequence was selected after processing all data sets 
with apertures of a wide range of diameters and choosing the one that
gave the lowest noise in the FT.
%{\bf i was choosing the best aperture based only on the lower mean amplitude in 
%the region 1.5--10\,mHz, not considering any maximum amplitude peak}
%using the amplitude of the primary peak
%for the signal and the mean of the amplitude spectrum outside the
%pulsation range for the noise.
The apertures tested ranged between 10 and 40 pixels in radius,
which corresponds to 1.9 and 7.6 arcseconds on the sky,
while the best results usually stayed within the range of 18 to 22 pixels
(3.4 to 4.2 arcseconds), but on nights with particularly poor seeing
the optimal radius could be as high as 30 pixels (5.7 arcseconds).

\begin{figure}[t]
\centering
\includegraphics[width=\hsize]{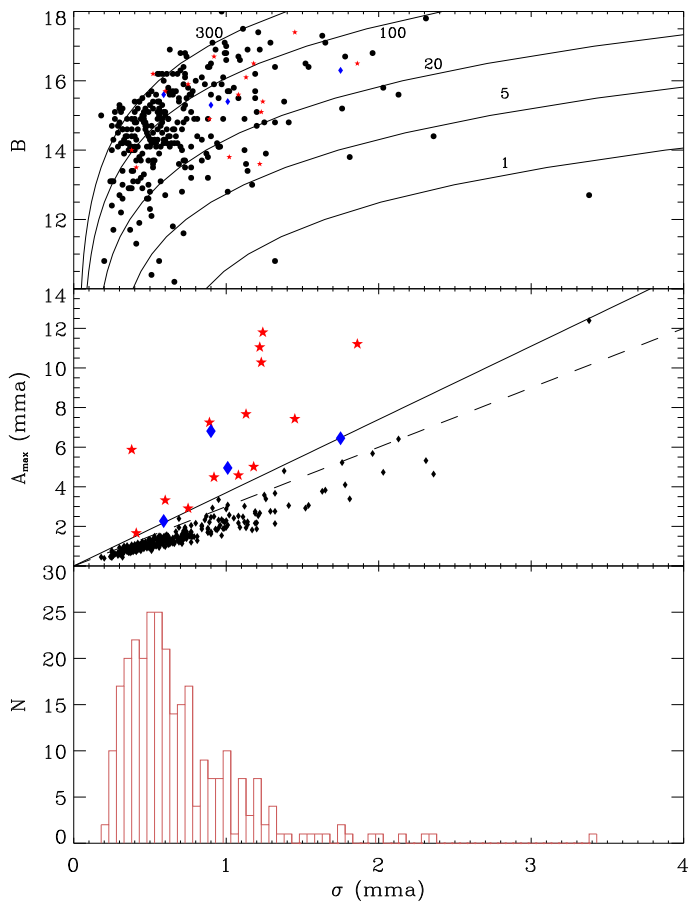}
\caption{Top panel: Photographic $B$ magnitudes of the targets
as function of $\sigma$.
The labeled curves indicate the theoretical noise limits for a typical
photometric observation with our system (1), and scaled to show the
limit achieved for a time-series of 5, 20, 100 and 300 measurements.
Middle: Maximum amplitude detected in the FT for each target,
as function of $\sigma$. The pulsating sdBs from this survey 
are marked with star symbols (published earlier) and
diamonds (presented in this paper).
The continuous line indicates the 3.7$\sigma$ level
(99\% confidence), and the dashed line indicates the 3.0$\sigma$ level.
Bottom: Histogram of the noise level achieved for the sample.}
\label{fig:signal}
\end{figure}

\section{Results}

% 285 + 20 (puls) + 4 (ecl+refl)  
A total of 309
stars deemed to lie in the predicted instability region for sdB stars
were checked for variability during 25 observing runs from 1999 to 2009.
We did not apply a specific cutoff for \teff\ and \logg\ since our
estimate of these parameters were preliminary and sometimes crude. Instead
we prioritised our targets by their distance from the centre of the
instability region at 34\,kK. In the later runs we shifted our focus more
towards the region where the hybrid pulsators are located.
Of all our targets, 20 were observed to pulsate
with frequencies corresponding to the V361\,Hya stars,
which corresponds to a $\sim 6.5\%$ ratio of fast pulsators/NOV sdBs.
A further four stars were found to have such strong reflection effects
that the slope was obvious even in our short light-curves. This fraction
of 1.3\%\ is only a lower limit for the sdB+dM population, as there may
be quite a few more objects with longer periods, which our survey for
short period pulsators was not sensitive to.
In fact, we know of one more reflection variable in our survey,
\object{BPS\,CS\,22169--1}, which we found to be constant in our
half-hour light-curve.
Since these results were not obtained at the {\sc not} as part of this
survey, they will be reported in a future paper.
% We recently reobserved this star with the Mercator
% telescope after \citet{edelmann05} presented an RV curve with a very low
% amplitude, indicating a system either with an extremely low mass companion
% or an sdB+dM seen almost pole on.  A reflection effect with an
% exceptionally low amplitude at the expected period was indeed found,
% and will be reported in a separate paper.
The 285 stars for which no obvious variability was detected
are presented in Table~\ref{tbl:limits}.

A different detection threshold was achieved for each target, depending
on its magnitude, the filter used for the observations ($B$ or $W$), the
length of the photometric time-series, and the weather conditions in
each particular case. In the top panel of Fig.~\ref{fig:signal} the
mean amplitude level in the FT ($\sigma$) obtained for the sample is shown 
as function of each star's magnitude. A general trend of larger $\sigma$
for fainter objects can be distinguished, although in most cases the
time-series sequence was stopped when the real-time FT showed no
significant peaks in the relevant region and the highest peaks reached
amplitudes $\sim$\,1\,mma.  Theoretical $\sigma$ levels as a function of
magnitude for light-curves with 1, 5, 20, 100 and 300 data points are also
plotted in this panel. They represent the contribution to $\sigma$ of both
the scintillation and photon shot noise, and thus should be understood as
the limiting accuracy in the FT obtainable from a time series with the same
number of data points.
The scintillation noise was computed following~\cite{young93}, and 
the photon shot noise was estimated using the signal-to-noise calculator
of the {\sc not}.  In both cases, a typical airmass $X=1.2$ and a typical
exposure time of 20\,s were used. Most points follow the predicted
trends for light-curves with between 20 and 300 points, and the outliers
are mostly due to poor weather conditions.

The middle panel of Fig.~\ref{fig:signal} shows, again as function of $\sigma$,
the maximum amplitude obtained in the FT for every observed target within the
1.5--10\,mHz frequency range.  The continuous line indicates the
3.7$\sigma$ (99\%) threshold, meaning that if $A_{\rm max}$ is 
above this line, the star is considered as pulsator. The dashed line
indicates the 3.0$\sigma$ (90\%) threshold, and stars whose $A_{\rm max}$
falls below this line are considered as NOV, according to our definition.
Fourteen of the twenty sdBVs discovered in this
survey are marked with stars (published earlier) and four with diamonds
(the new ones presented here); the remaining two pulsators,
\object{EP\,Psc} and \object{QQ\,Vir}, have such large pulsation amplitudes
that they lie beyond the range shown in the figure.
Note that we have used only the discovery data for these plots, so the
lowest amplitude pulsator, \object{LM\,Dra}, and the one on the
99\%\ confidence line at $\sigma$\,=\,0.75, \object{PG\,1657+416},
have been confirmed in follow-up runs with higher confidence levels.
The objects between the 3.0 and 3.7\,$\sigma$ confidence lines are
discussed below, and the pulsators marked with diamonds will be discussed
in Sect.~\ref{sect:pulsators}.

The bottom panel of Fig.~\ref{fig:signal} represents an histogram of the mean
amplitude level achieved along the survey. For most of the targets, the noise
level in the FT was below 1\,mma (239 targets = $83\%$), and in 50\% of the
cases a $\sigma < 0.6$\,mma was obtained.

\subsection{Pulsator candidates}\label{sect:pcand}

Ten objects in Table~\ref{tbl:limits} have amplitude peaks between 3.0
and 3.7\,$\sigma$, higher than our NOV requirement.
They must therefore still be considered pulsator candidates.
Three of these have a $\sigma$ of 0.33\,mma or less, and have been ignored.
The others deserve particular mention.

One object, \object{PG\,1248+164} has such a high $\sigma$ that it
is off the scale in Fig.~\ref{fig:signal}.
The object comes from the sample
of \citet{maxted01} and deserves further observations.
The object on the 99\% confidence line at $\sigma$\,=\,3.4 is
\object{PG\,1725+252}. Normally it would have been reobserved, but as it was
realised that this object is also in the survey of \citet{billeres02} with
a comfortable 0.03\,\% limit, no follow-up was performed.
The object just below the 99\%\ limit at $\sigma$\,=\,0.7 is
\object{KPD\,2215+5037}, observed during run N5.
This star was also part of the survey of \citet{billeres02}
with a limit of 0.2\,\%. We note a rising trend in our light-curve on
the 1\%\ level over the 20 minutes covered by our observations, which could
indicate that the object is a long period variable.

The three objects between the 3\,$\sigma$ and 3.7\,$\sigma$ confidence lines
at $\sim$1$\sigma$ are \object{PG\,1313+132}, \object{PG\,1722+286} and
\object{J1351+0234}. \object{PG\,1313+132} shows variations at low frequencies,
consistent with $g$-mode pulsations, which is likely considering its low
temperature \citep[\teff\,=\,25.6\,kK, \logg\,=5.41;][]{saffer94},
but uncertain as the observations were done at relatively high airmass.
If we compute the noise level only for frequencies higher than 5\,mHz we
get $\sigma$\,=\,0.53 and the highest peak is well below three times this value.
The light-curve of \object{PG\,1722+286} was obtained during the problematic
N1 run.  An attempt was made to reobserve it during T2, but poor weather
resulted in an even
worse light-curve. After a limit on pulsation of 0.08\%\ was presented
by \citet{billeres02}, no further attempts were made to observe this target.
The last of the three,
\object{J1351+0234}, is just noisy due to its low brightness ($g'$\,=\,17.1).

The three objects that lie on the 3$\sigma$ confidence line around
$\sigma$\,=\,2 are \object{HS\,1813+7247}, \object{EGGR\,491}, and
\object{HS\,2029+0301} (only the last of these is actually above the
3\,$\sigma$ line, but we will discuss all three).
\object{HS\,1813+7247} has a slope in the light-curve, but it is not consistent
between the different reference stars, and so is most likely due to differential
extinction as the $W$-band filter was used. \object{EGGR\,491} is another faint
target from the SDSS sample, so the light-curve is just noisy, with no obvious
features. \object{HS\,2029+0301} is also from the N1 photoelectric run, and
has not been reobserved. Further observations would be required to give a more
useful limit on any pulsational behaviour of this object.

\section{New pulsators}\label{sect:pulsators}

The four pulsators found in our sample and not published elsewhere
are \object{HE\,2151--1001}, \object{HS\,2125+1105}, \object{PG\,1033+201},
and \object{HE\,1450--0957}. The details of the discovery observations
as well as follow-up observations on the first two done with the {\sc not}
are provided here.  Table~\ref{tbl:newsdBVs} includes
the detailed log of observations for all these new sdBVs.

\begin{figure}[t]
\includegraphics[width=\hsize]{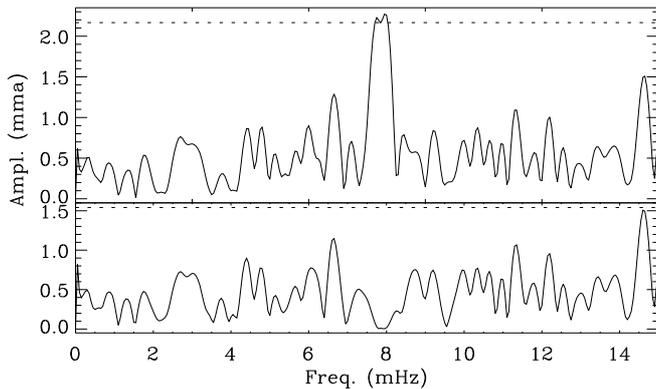}
\caption{Amplitude spectrum of HE\,2151--1001. 
The lower panel shows the residual spectrum after prewhitening
the two frequencies listed in Table~\ref{tbl:freqs}
from the light-curve.
The dashed line indicates 3.7 times the noise level.}
\label{fig:he2151}
\end{figure}

\begin{figure}[t]
\includegraphics[width=\hsize]{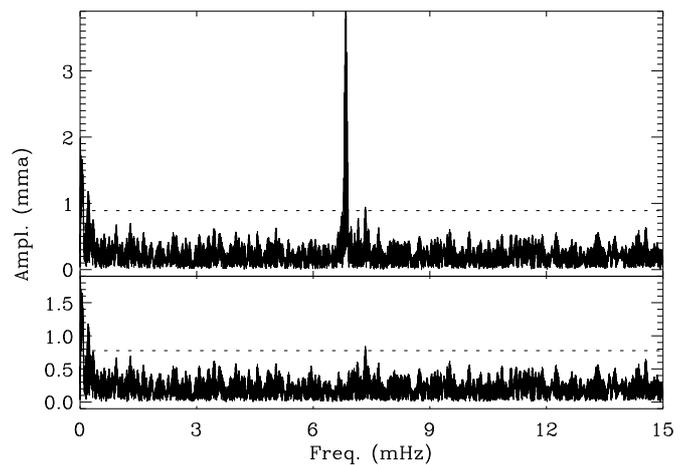}
\caption{Amplitude spectrum of HS\,2125+1105. The lower panel shows residual spectrum after 
subtracting one frequency from the original light-curve.
The dashed line indicates 3.7 times the noise level.
The peak at 7.3\,mHz is just above the 3.7\,$\sigma$ level if the low
frequency domain is excluded when estimating $\sigma$.
}
\label{fig:hs2125}
\end{figure}

For all the pulsators we derived the frequency content of the light-curves
by performing a non-linear least-squares fit to a sine function, using the
highest amplitude frequency in the amplitude spectrum (top panels
of Fig.~\ref{fig:he2151}, \ref{fig:hs2125}, \ref{fig:pg1033} and
\ref{fig:he1450}). This fit was then subtracted from the light-curve
to compute the residual amplitude spectrum (bottom panels).
Notice that the noise level is recomputed after each prewhitening step,
so that the 3.7$\sigma$ level lines in the figures drop after removing
the established signal.
If there was still any peak left above the significance threshold,
the {\em original} light-curve was fitted to the sum of two sine
functions with the established frequencies.
Table~\ref{tbl:freqs} lists the best fitting frequencies and amplitudes,
with the associated errors from the least-squares procedure.

\subsection{HE\,2151--1001}

This target was observed for $\sim$40\,min during the run S2.
Its amplitude spectrum is shown 
in Fig.~\ref{fig:he2151}, where some signal above the threshold can
be distinguished at $\sim$8\,mHz. 
This frequency range is typical for a pulsating sdB with
these~\teff-\logg\ parameters (see Table~\ref{tbl:sdbvars}). 

\begin{table}[b]
\caption[]{Details from the discovery runs for the four new sdBVs.}
\label{tbl:newsdBVs}
\begin{center}
\begin{tabular}{lcrccc} \hline\hline
Target        &Run &$N_p$& $\Delta t$ & $\sigma$ & $A_\mathrm{max}$\\
              &ID  &    & s  & mma  & mma \\ \hline\\[-2.5mm]
HE\,2151--1001& S2 &120 & 30 & 0.58 & 2.20 \\%(NOT7.table=NOT_May05)
              & S3 & 73 & 30 & 1.05 & 2.35 \\% (NOTOct07.table)
HS\,2125+1105 & S2 & 67 & 35 & 1.75 & 6.45 \\%(NOT7.table=NOT_May05)
              & ST & 95 & 30 & 0.95 & 3.56 \\
              & S3 &116 & 42 & 0.72 & 3.42 \\%(NOT_Oct07=S3)
              & S3 &433 & 30 & 0.40 & 4.35 \\%(NOT_Oct07=S3)
              & S3 &541 & 30 & 0.33 & 3.84 \\%(NOT_Oct07=S3)
              & S4 & 89 & 28 & 1.17 & 4.76 \\%(NOT_Sept08=S4)
PG\,1033+201  & T7 &118 & 20 & 1.01 & 4.95 \\
HE\,1450--0957& S6 &125 & 30 & 0.90 & 6.81 \\%(NOT_EN09.table)
\hline\\[-6.0mm]
\end{tabular}\end{center}
`Run ID' refers to the identifiers listed in Table~\ref{tbl:obstime},
$N_p$ is the number of points for the specific sequence, $\Delta t$ is the
sampling time interval, $\sigma$ is the mean noise level in the FT,
and $A_\mathrm{max}$ is the amplitude of the highest peak.
\end{table}

Our non-linear least-squares frequency analysis revealed two low amplitude
(3\,mma) peaks, as listed in
Table~\ref{tbl:freqs}. Note that the frequency distance between the two
peaks ($\sim$140\,$\mu$Hz) is within the 
frequency resolution ($\sim$325\,$\mu$Hz).
Another short sequence on \object{HE\,2151--1001} was obtained during S3
(see Table~\ref{tbl:newsdBVs}), but the noise level achieved was not sufficient 
to detect the oscillations above the significance threshold.

This target was observed spectroscopically as a white dwarf candidate by
the {\sc spy} survey. The spectrum was analysed by
\citet{lisker05}, who found no spectral indication of a companion,
but noted that it has a peculiar H$\alpha$ profile.
The \twomass\ photometry ($J$\,=\,16.53, $H$\,=\,16.37, $K$\,=\,15.91)
indicates a main sequence companion, but this is uncertain as the values
are close to the faint limit of the \twomass\ survey.
The {\sc nomad} catalogue
\citep[Naval Observatory Merged Astrometric Dataset,][]{zacharias05}
provides reasonably good photographic magnitudes 
($B$\,=\,15.25, $V$\,=\,15.57, $R$\,=\,16.09),
and the UV magnitudes ($FUV$\,=\,14.55, $NUV$\,=14.87) from the
{\sc galex} satellite \citep{GALEX} all indicate colours
typical for a hot subdwarf.

\subsection{HS\,2125+1105}

The faint ($B$\,=\,16.3) \object{HS\,2125+1105} was first observed during run
S2.  A 6.4\,mma peak was detected at 6.89\,mHz, although its significance
level was exactly 3.7 times the noise level (see Table~\ref{tbl:newsdBVs}).
A year later, a slightly longer light-curve with better photometric conditions
was obtained during run ST, which reduced the noise level to 0.95\,mma.
However, a peak detected at the same frequency also dropped its amplitude
to 3.6\,mma, again at exactly the 3.7 times the noise level.
During run S3, the object was observed on each of the
three nights as indicated in Table~\ref{tbl:newsdBVs}, and the good
conditions finally allowed us to get reliable confirmation of the main
peak.
The amplitude spectrum obtained from the combined light-curve is shown 
in the upper panel of Fig.~\ref{fig:hs2125}.
In Table~\ref{tbl:freqs} the results of a frequency 
analysis are listed. A 4\,mma peak at 6.83\,mHz is detected,
but this time at a reassuringly high S/N. In this sequence a second peak at
7.34\,mHz was also detected, as listed in Table~\ref{tbl:freqs}.
The bottom panel of Fig.~\ref{fig:hs2125}
shows the residual amplitude spectrum after prewhitening the main frequency
from the light-curve.

In order to check the stability of this mode, an additional short sequence was
taken in $S4$. The same $f_1$ was found, this time with a 4.7\,mma amplitude.

This HS star was not part of the original sample of sdB stars from
\citet{EdelmannPhD}, but was observed as part of the {\sc spy} survey and was
analysed by \citet{lisker05}.
The star has reliable optical photometry (but no spectrum)
in the SDSS survey ($u'$\,=\,16.13, $g'$\,=\,16.36, $r'$\,=\,16.78,
$i'$\,=\,17.09, $z'$\,=\,17.33).
There is no indication of a main sequence companion from the spectroscopy or
SDSS photometry. The object is well below the faint limit of the \twomass\
survey, so no IR colours are available. 
The {\sc galex} photometry is FUV\,=15.6, NUV\,=\,15.8.

\begin{figure}[t]
\includegraphics[width=\hsize]{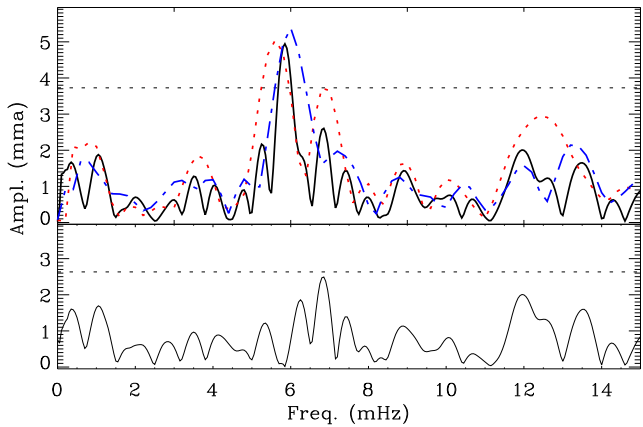}
\caption{Amplitude spectrum of PG\,1033+201.
The lower panel shows the residual spectrum after prewhitening
the two frequencies listed in Table~\ref{tbl:freqs} from the light-curve.
We have also plotted the amplitude spectrum of the first half (red dashed curve) and
second half (blue short dash, long dash curve), to demonstrate that
the main peak is present with the same amplitude in both halves of the
dataset.  The dashed line indicates 3.7 times the noise level.
}
\label{fig:pg1033}
\end{figure}

\begin{table}[b]
\caption[]{Pulsation properties for the four new pulsators.}
\label{tbl:freqs}
\begin{center}
\begin{tabular}{lcccr} \hline\hline
Name & Id & Frequency & Amplitude & S/N\\
 & & [$\mu$Hz] & [mma] & \\\hline\\[-2.5mm]
HE\,2151--1001
 & $f_1$ & 7931\,$\pm$\,63 & 3.04\,$\pm$\,1.81  & 6.5\\
 & $f_2$ & 7791\,$\pm$\,64 & 3.00\,$\pm$\,1.80  & 6.5\\\hline\\[-2.5mm]
HS\,2125+1105
 & $f_1$ & 6835.66\,$\pm$\,0.07 & 4.05\,$\pm$\,0.19 & 19.6\\
 & $f_2$ & 7342.35\,$\pm$\,0.35 & 0.85\,$\pm$\,0.19 &  4.1\\\hline\\[-2.5mm]
PG\,1033+201
 & $f_1$ & 5850\,$\pm$\,27 & 4.9\,$\pm$\,0.6 &  8.1\\
 & $f_2$ & 6837\,$\pm$\,53 & 2.5\,$\pm$\,0.6 &  4.5\\\hline\\[-2.5mm]
HE\,1450--0957
 & $f_1$ & 7202\,$\pm$\,8  & 6.9\,$\pm$\,0.4 & 13.7\\
 & $f_2$ & 8475\,$\pm$\,21 & 2.5\,$\pm$\,0.4 & 5.2\\\hline\\[-6.0mm]
\end{tabular}\end{center}
Pulsation frequencies and amplitudes are given
as derived by non-linear least-square fitting to the light-curves.
The S/N is computed after prewhitening the detected frequencies from the
signal.
\end{table}

\begin{figure}[t]
\includegraphics[width=\hsize]{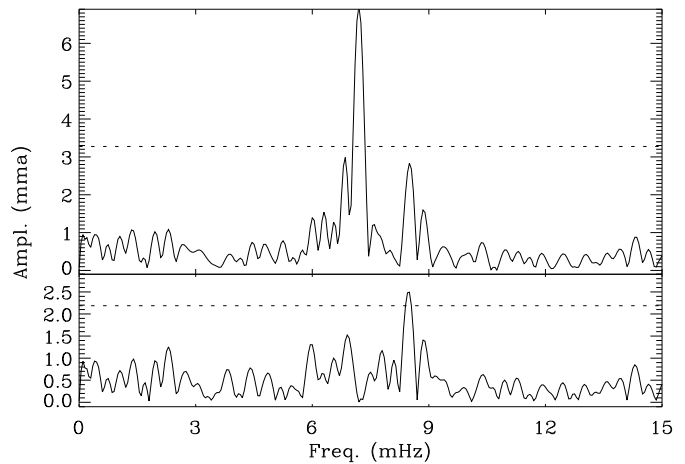}
\caption{Amplitude spectrum of HE\,1450--0957 (top) and residual amplitude spectrum
after subtracting the main peak (bottom), with 3.7$\sigma$ noise levels
(dashed lines).
}
\label{fig:he1450}
\end{figure}

\subsection{PG\,1033+201}

\object{PG\,1033+201} was one among 15 targets observed 
with the $B$-band filter during the technical night T7.
During the run we had some problems with the observing system, which resulted
in some frames in the photometric sequence being dropped. For this reason
the real-time processing program was unable to process the entire sequence.
Due to these problems we stayed on target almost 40 minutes, which is
considerably longer than our standard sequence on such bright targets.
Eventually, we
terminated the observations in order to reinitialise the CCD system, but
we never returned to reobserve this star. During the careful reprocessing
of the photometry for this paper, we excluded two lost and one bad data point
from the 118 frame sequence, and performed a regular frequency analysis of
the cleaned light-curve.
The amplitude spectrum is shown in the top panel of
Fig.~\ref{fig:pg1033} as a continuous line. A clear peak at 5.8\,mHz is
well above the 3.7$\sigma$ (horizontal dashed) line. We divided the
light-curve into two equal halves, and we can see from the corresponding
spectra (also shown in Fig.~\ref{fig:pg1033} as red dotted and blue
dash-dotted curves), that the main period is present in both halves with
the same amplitude. A second frequency at 6.8\,mHz is significant only in the first
part, indicating that it is either spurious or a combination of frequencies
suffering cancellation from beating within our short sequences.
This exercise demonstrates how short-period pulsations can be detected with
reasonable confidence even in short ($\sim$20\,m) light-curves, as long as
one acquires adequate signal-to-noise in the individual data-points.
The bottom panel displays the residual amplitude spectrum after subtracting
the frequency at 5.8\,mHz. The remaining peak at $\sim$6.8\,mHz becomes
significant when the noise is computed in the prewhitened spectrum
(see Table~\ref{tbl:freqs}).
We conclude that \object{PG\,1033+201} appears to be
a typical short-period sdB pulsator with a main pulsation amplitude of
5\,mma, with likely further frequencies at the 2\,mma level. 

This object was included in our sample based on the decomposition by
\citet{allard94}, who estimated the system to be an sdB+F9 composite
from their Cousins BVRI photometry ($V$\,=\,15.67, $B$--$V$\,=\,--0.17,
$V$--$R$\,=\,--0.01, $R$--$I$\,=\,+0.06), and provided a colour temperature
for the primary of 31.5\,kK. 
The object occurs in the SDSS survey as J103638.93+195202.2, and
has photometry ($u'$\,=\,15.14, $g'$\,=\,15.39, $r'$\,=\,16.80,
$i'$\,=\,16.01, $z'$\,=\,16.21) but no spectroscopy, which prevents
us from acquiring a better temperature and a gravity estimate.
The \twomass\ IR photometry clearly supports the notion of a
main-sequence companion \citep[$J$\,=\,15.46, $H$\,=\,15.16, $K$\,=\,15.13,
i.e.~$J-H$\,$>$\,+0.3, see][]{reed04}.

\subsection{HE\,1450--0957}

\object{HE\,1450--0957} was one of the three targets surveyed during the
final run of the programme (S6).
A rather long light-curve was obtained (1h5min), since its oscillations were
already clear as the data were processed as they arrived.
Its amplitude spectrum (upper panel of Fig.~\ref{fig:he1450})
reveals the multimode behaviour of this target.
At almost 7\,mma, the main peak at 7.2\,mHz has the highest amplitude of
the four new pulsators presented here.
A significant second peak is also detected in the amplitude spectrum of
\object{HE\,1450--0957} (see bottom panel of Fig. 5),
but more low amplitude modes are likely to remain in the range 6\,--\,9\,mHz.
The frequencies derived from this discovery light-curve are also listed
in Table~\ref{tbl:freqs}. 

This object came into our sample from the HE stars surveyed by {\sc spy}
\citep{lisker05}. It is included in the Edinburgh-Cape survey as 
\object{EC\,14507--0957} where reliable photometry can be found
($V$\,=\,15.27, $U$--$B$\,=\,$-$1.04, $B$--$V$\,=\,$-$0.21).
As noted by \citet{reed04}, its \twomass\ magnitudes
($J$\,=\,15.58, $H$\,=\,15.41, $K$\,=\,15.36) are indicative of a
main sequence companion, but this companion must be very weak in the
optical as \citet{lisker05} do not detect any trace of it in
their VLT/\uves\ spectrum.

\section{Summary and discussion}

The {\sc not} search programme for pulsating sdB stars has been a great
success,
contributing twenty new pulsators to the known population of short period
sdBVs. Together with the two new pulsators presented by \citet{kilkenny09},
another two recently found by \citet{barlow09,barlow10},
and \object{HE\,0230--4323}\footnote{Originally reported as
``an unusual hot subdwarf pulsator'' by \citet{koen07} but now confirmed to
be a more regular sdBV by Kilkenny (priv.comm)},
our latest tally brings the total up to 49.
Our effort to implement an efficient system for CCD photometry at
the {\sc not} has certainly been fruitful, producing on average one new
pulsator for every three nights of observations. This is a substantially
higher efficiency than any other group has reported.
% Our efforts to
% implement an efficient fast photometry system on the instruments
% at the {\sc not} has certainly paid off, although it remains an
% add-on that is not supported by the regular observing system.

\begin{figure*}[t!]
\centering
\includegraphics[width=9.8cm,angle=-90,clip]{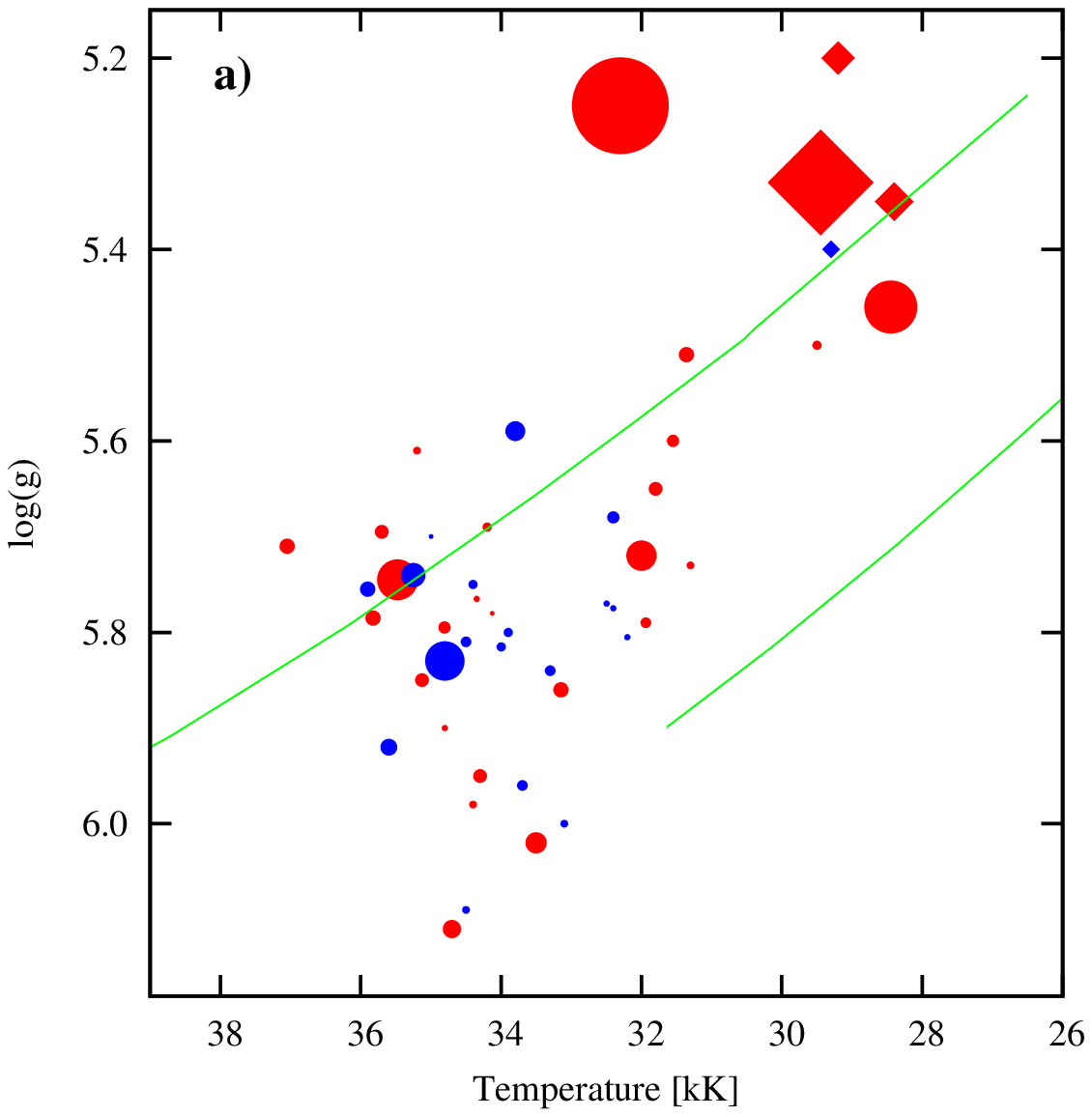}
\includegraphics[width=9.8cm,angle=-90,clip]{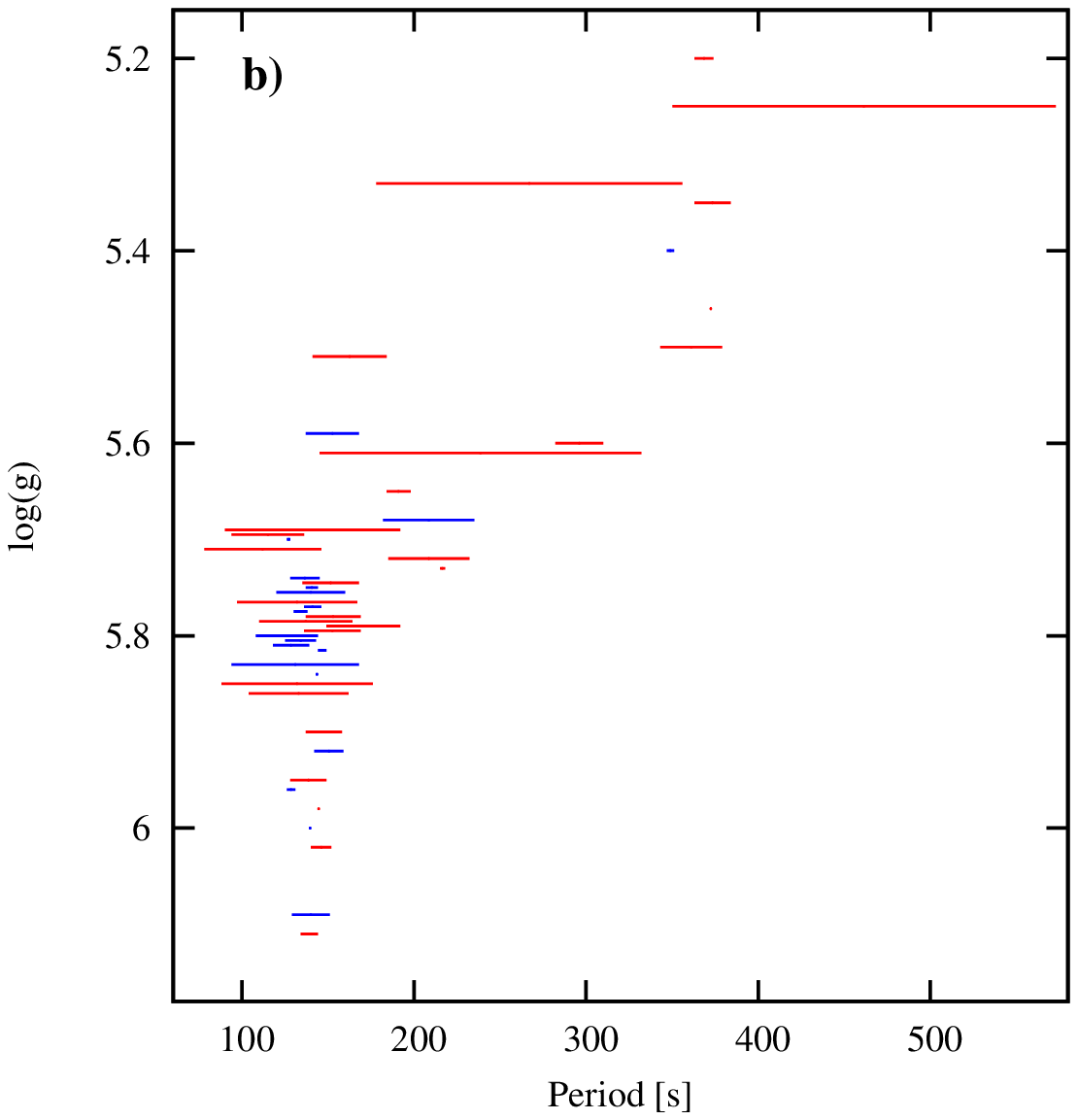}
\caption{a) \teff--\logg\ diagram for the objects in Table~\ref{tbl:sdbvars}.
Filled circles indicate V361\,Hya type pulsators and diamonds indicate hybrid
DW\,Lyn type pulsators.
The size of the symbols is proportional to the pulsation amplitude,
$A_\mathrm{max}$.
The nineteen sdBVs from this survey that we have \teff\ and \logg\ for
are shown in blue, and short period pulsators from other surveys in red.
The green lines indicate the canonical zero age EHB (lower line) and
terminal age EHB for an 0.47\,\msol\ core \citep[from][]{kawaler05}
with envelope mass increasing from bottom to top.
The ZAEHB starts at a \logg\ of 5.9
with an envelope mass fraction of only 0.02\%, and the objects that appear to
lie on the extension of this line cannot be explained by the canonical models.
b) $P$--\logg\ diagram for the same pulsators.
The bars indicate the range of detected pulsation periods in any particular
star, excluding harmonics and $g$-modes.
The \logg\ values have been shifted by up to $\pm0.04$~dex
to avoid overlaps.
}
\label{fig:tpg}
\end{figure*}

\subsection{Follow-up}
Many of our pulsators have already received detailed follow-up.
Most effort has been put in on \object{V391\,Peg}, which was the subject of
several campaigns, and its pulsation periods are still being monitored regularly
\citep{silvotti07}. The original aim of this monitoring was to
detect evolutionary period changes, but the periodic variation that was
found is more likely to be caused by an orbiting planet inducing
period variations through the light travel time effect.
Several other stars in our sample are now regularly monitored for
variations in the pulsation period under the {\sc exotime} project
\citep{exotime}. As a substantial time base of these observations develops,
we expect to detect more planet-hosting hot subdwarf stars.

\object{QQ\,Vir} was the first sdBV for which
spectroscopic line-profile variations were used to constrain its main
pulsation mode \citep{telting04}, and was the subject of an extensive
photometric campaign in 2003 \citep{silvotti06} revealing 15 pulsation
frequencies. The detailed frequency spectrum from this campaign allowed
\citet{charpinet06} to obtain an asteroseismic solution using the forward method.
\object{QQ\,Vir} also became the first star from
our sample to be targeted with high-resolution spectroscopy with
the VLT in 2008 \citep{telting10}.

\object{LM\,Dra} was followed extensively by \citet{reed07b} with
multisite observations between 2003 and 2004,
and in spite of its extremely low pulsation amplitudes they
detected six pulsation periods with amplitudes between 1.0 and 2.2\,mma.
\object{V429\,And} and \object{V1636\,Ori} were followed by \citet{reed07a}
in Nov--Dec\,2005, revealing a rich frequency spectrum with fourteen significant
peaks between 0.45 and 4.35\,mma in \object{V429\,And}, and a simpler
spectrum with only three significant peaks between 0.8 and 11.1\,mma
in \object{V1636\,Ori}. \citet{reed07a} also report that \object{LS\,Dra}
appears to be monoperiodic, but with a highly variable amplitude,
changing from a maximum of 5.26 to a minimum of 0.87\,mma
over their campaign spanning 47\,days.  They further confirm three of the four
pulsation periods in \object{V387\,Peg}.

\object{J1717+5805} was targeted with high-speed simultaneous 3-channel
photometry using {\sc ultracam} on the WHT, as reported by \citet{aerts06},
resolving the broad peak described in \citetalias{solheim04} into two
separate frequencies.

\begin{table}[t]
\caption[]{Comparison with the results of other surveys for pulsating
sdBs (upper part) and our own survey divided into subsamples (lower part).
}
\label{tbl:stat}
\begin{center}
\begin{tabular}{lrrrrl} \hline\hline
Sample   & Size & \multicolumn{2}{c}{Pulsators} & Fract. & Preselection \\
         & \#  & S & C & \%  & on \teff \\\hline\\[-2.8mm]
South Africa & $\sim$1200 & 13 & 7 & 1.7 & No \\
Billeres & 74 & 0 & 4 &  5.4 & Yes \\
Dreizler & 12 & 1 & 0 &  8   & Yes \\
\hline\\[-2.8mm]
HS       & 80 & 9 & 0 & 10   & Yes, --comp. \\
HE       & 23 & 0 & 0 &  0   & Yes \\
SPY      & 27 & 3 & 0 & 11   & Yes \\
SDSS     & 40 & 0 & 5 & 12.5 & Yes \\
BG       & 11 & 1 & 0 & 9    & Yes \\
Saffer   & 24 & 1 & 0 & 4    & Yes \\
Maxted   & 24 & 0 & 0 & 0    & Yes \\
colour   & $\sim$50 & 0 & 2 & 4 & Poor \\ \hline\\[-6.0mm]
\end{tabular}\end{center}
'Size' indicates the number of targets checked for variability in each
sample. The number of pulsators discovered are separated into
apparently single (S) and spectroscopic composites (C),
but note that binaries with invisible (WD or dM) companions are included
in the column for single pulsators.
We also specify if the sample was preselected to contain only subdwarfs
in the instability region or not.
Note that a few of the pulsators are present in more than one sample, and
that we keep the HS and HE stars observed only as part of the {\sc spy} survey
apart from the HS and HE samples.
\end{table}

\begin{table*}[th!]
\caption[]{All short period sdBV stars published to date.}
\label{tbl:sdbvars}
\begin{center}
\begin{tabular}{llccccrrrrl} \hline\hline
Name               & Survey name    & $m_V$& \teff & \logg& \logy & $C$ &
                     $P_\mathrm{range}$ & $A_\mathrm{max}$ & $N_P$ & References \\
 & & mag & kK & dex & dex & & s & mma & & \\ \hline\\[-2.8mm]
\object{V361\,Hya}   & EC\,14026--2647 & 15.3 & 34.7 & 6.10 &  ...   &  G2   & 134--144 &  12 &  2 & $^{1,4}$ \\ % Chk
\object{EO\,Cet}     & PB\,8783        & 12.3 & 35.7 & 5.70 &  ...   &  F0   &  94--136 &   9 &  7 & $^{2,4}$ \\ % Chk
\object{UX\,Sex}     & EC\,10228--0905 & 15.9 & 33.5 & 6.00 &  ...   &  G2   & 140--152 &  14 &  3 & $^{3,4}$ \\ % Chk
\object{V4640\,Sgr}  & EC\,20117--4014 & 12.5 & 34.8 & 5.87 &  ...   &  F5   & 137--158 &   4 &  3 & $^4$ \\ % Chk
\object{UY\,Sex}     & PG\,1047+003    & 13.5 & 33.2 & 5.80 & --2.00 &$^{ab}$& 104--175 &  10 & 18 & $^{5,k1}$ \\ % Chk
\object{NY\,Vir}     & PG\,1336--018   & 13.5 & 31.3 & 5.60 & --2.93 &  M5   &  97--205 &  10 & 28 & $^{6,k2,v2}$ \\ % Chk
\object{V2203\,Cyg}  & KPD\,2109+4401  & 13.4 & 31.8 & 5.79 & --2.23 &$^{a}$ & 182--213 &   9 &  8 & $^{7,z1,h1}$ \\ % Chk
\object{V338\,Ser}   & PG\,1605+072    & 12.8 & 32.3 & 5.25 & --2.53 &$^{a}$ & 350--573 &  64 & 50 & $^{8,12}$ \\ % Chk
\object{KL\,UMa}     & Feige\,48       & 13.5 & 29.5 & 5.50 & --2.93 &  WD   & 343--379 &   6 &  4 & $^{9,h1,o2}$ \\ % Chk
\object{DT\,Lyn}     & PG\,0911+456    & 14.6 & 31.9 & 5.77 & --2.55 &$^{a}$ & 149--192 &   7 &  7 & $^{10,r5}$ \\ % Chk
\object{KY\,UMa}     & PG\,1219+534    & 13.2 & 34.3 & 5.95 & --1.50 &$^{ab}$& 128--149 &   9 &  4 & $^{10,h1}$ \\ % ?
\object{AQ\,Col}     & EC\,05217--3914 & 15.6 & 31.3 & 5.76 & --3.00 & ...   & 215--218 &   8 &  8 & $^{11,r3}$ \\ % Chk
\object{V1405\,Ori}  & KUV\,04421+1416 & 15.1 & 32.0 & 5.72 & --2.50 &  dM   & 185--232 &  20 &  7 & $^{11,r4}$ \\ % Chk
\object{V2214\,Cyg}  & KPD\,1930+2752  & 13.8 & 35.2 & 5.61 & --1.50 &  WD   & 145--332 &   5 & 44 & $^{13,g1}$ \\ % Chk
\bf\object{LM\,Dra}      & PG\,1618+563B   & 13.5 & 33.9 & 5.80 & --1.60 &F3$^c$ & 108--144 &   6 &  2 & $^{14,r3}$ \\ % Chk
\bf\object{DV\,Lyn}      & HS\,0815+4243   & 16.1 & 33.7 & 5.95 & --2.10 & ...   & 126--131 &   7 &  2 & $^{15}$ \\ % Chk
\bf\object{V384\,Peg}    & HS\,2149+0847   & 16.5 & 35.6 & 5.90 & --1.80 & ...   & 142--159 &  11 &  5 & $^{15}$ \\ % Chk
\bf\em\object{V391\,Peg} & HS\,2201+2610   & 13.6 & 29.3 & 5.40 & --3.00 &  Pl   & 347--351 &  10 &  3 & $^{15,s1}$ \\ % Chk
\bf\object{LS\,Dra}      & HS\,1824+5745   & 15.6 & 33.1 & 6.00 & --1.52 & ...   &      139 &   5 &  1 & $^{16,r1}$ \\ % Chk
\bf\object{V387\,Peg}    & HS\,2151+0857   & 16.5 & 34.5 & 6.10 & --1.37 & F--K  & 129--151 &   4 &  5 & $^{16,r1}$ \\ % Chk
\bf\object{V429\,And}    & HS\,0039+4302   & 15.1 & 32.4 & 5.70 & --2.20 &$^{a}$ & 182--235 &   8 &  4 & $^{16}$ \\ % Chk
\bf\object{V1636\,Ori}   & HS\,0444+0458   & 15.4 & 33.8 & 5.60 & --1.85 &$^{a}$ & 137--168 &  13 &  2 & $^{16}$ \\ % Chk
\object{EK\,Psc}         & PG\,0014+067    & 16.3 & 34.1 & 5.77 & --1.68 &$^{b}$ & 137--169 &   3 & 13 & $^{17,c1,v1}$ \\%Chk
\bf\object{EP\,Psc}      & PG\,2303+019    & 16.2 & 35.3 & 5.74 & --1.70 &  MS?  & 128--145 &  16 &  3 & $^{18}$ \\ %Chk
\bf\object{QQ\,Vir}      & PG\,1325+101    & 13.8 & 34.8 & 5.81 & --1.65 &$^{ab}$&  94--168 &  26 & 14 & $^{18,t1,s2}$\\%Chk
\em\object{DW\,Lyn}      & HS\,0702+6043   & 14.7 & 28.4 & 5.35 & --2.70 &$^{a}$ & 363--384 &  22 &  2 & $^{19,s3}$ \\%Chk
\object{V1078\,Her}      & PG\,1613+426    & 14.1 & 34.4 & 5.97 & --1.65 &$^{a}$ &      144 &   5 &  1 & $^{20}$ \\ % Chk
\object{PG\,0048+091}    & ...             & 14.3 & 34.2 & 5.69 & --3.00 &  F5   &  90--192 &   6 & 28 & $^{21,r3,\ddag}$\\%Chk
\object{PG\,0154+182}    & ...             & 15.3 & 35.8 & 5.80 & --1.67 & G--K  & 110--164 &  10 &  6 & $^{21,r1,\ddag}$\\%Chk
\bf\object{J1717+5805}   & ...             & 16.7 & 34.4 & 5.75 & --1.80 &  G9   & 137--144 &   6 &  2 & $^{22,\dag}$\\% Chk
\em\object{Balloon\,090100001}& ...        & 11.8 & 29.4 & 5.33 & --2.54 &$^{ab}$& 178--356 &  60 &  3 & $^{23,o1}$ \\ % Chk
\bf\object{PG\,1419+081} & ...             & 14.9 & 33.3 & 5.85 & --1.80 &  G5   &      143 &   7 &  1 & $^{24,\dag}$ \\ % Chk
\bf\object{J1445+0002}   & ...             & 17.4 & 35.9 & 5.75 & --1.60 &  G0   & 120--142 &   8 &  3 & $^{24,\dag}$ \\ % Chk
\bf\object{J1642+4252}   & ...             & 15.7 & 32.4 & 5.80 & --2.00 &  G0   & 130--138 &   3 &  2 & $^{24,\dag}$ \\ % Chk
\object{EC\,09582--1137} & PG\,0958--116   & 15.2 & 34.8 & 5.79 & --1.68 &$^{a}$ & 136--169 &   8 &  5 & $^{25,r6}$\\%Chk
\object{EC\,11583--2708} & ...             & 14.4 &  ... & ...  &  ...   &  MS   & 114--149 &   3 &  4 & $^{25}$ \\ % Chk
\object{EC\,20338--1925} &BPS\,CS\,22880--18& 15.6& 35.5 & 5.75 & --1.71 &$^{a}$ & 135--168 &  27 &  5 & $^{25,\ddag}$\\%Chk
\em\object{RAT\,J0455+1305}& ...           & 17.2 & 29.2 & 5.20 &  ...   & ...   & 363--374 &  19 &  2 & $^{26,b1}$ \\ % Chk
\bf\object{PG\,1657+416}   & ...           & 15.9 & 32.2 & 5.80 & --1.60 &  G5   & 125--143 &   4 &  5 & $^{27,\dag}$ \\ % Chk
\bf\object{2M0415+0154}    & ...           & 14.0 & 34.0 & 5.80 & --1.60 &$^{ab}$& 144--149 &   6 &  3 & $^{28}$ \\ % Chk
\object{EC\,01541--1409}   & ...           & 12.2 & 37.1 & 5.71 & --1.21 &$^{a}$ &  78--146 &  10 &  6 & $^{29,\ddag}$ \\ % Chk
\object{EC\,22221--3152}   & ...           & 13.4 & 35.1 & 5.85 & --1.47 &$^{a}$ &  84--176 &   9 & 10 & $^{29,\ddag}$ \\ % Chk
\object{JL\,166}           & ...           & 15.0 & 34.4 & 5.75 & --0.80 &  MS   &  97--167 &   4 & 10 & $^{30}$ \\ % Chk
\object{CS\,1246}   & CS\,124636.2--631549 & 14.6 & 28.5 & 5.46 & --2.00 &$^{b}$ &      372 &  35 &  1 & $^{31}$ \\
\object{HE\,0230--4323} & ...              & 13.8 & 31.6 & 5.60 & --2.58 &  dM   & 282--310 &   8 &  3 & $^{32,l1}$\\%Chk
\bf\object{HE\,2151--1001} & ...           & 15.6 & 35.0 & 5.70 & --1.60 &  MS   & 126--128 &   3 &  2 & $^{\dag,l1}$\\%Chk
\bf\object{HS\,2125+1105}  & ...           & 16.3 & 32.5 & 5.76 & --1.86 &$^{a}$ & 136--146 &   4 &  1 & $^{\dag,l1}$\\%Chk
\bf\object{PG\,1033+201}   & ...           & 15.4 & 31.5 & ...  &  ...   &  F9   & 146--171 &   5 &  2 & $^{\dag,a1}$ \\%Chk
\bf\object{HE\,1450--0957} &EC\,14507--0957& 15.3 & 34.6 & 5.79 & --1.29 &  MS   & 118--139 &   7 &  3 & $^{\dag,l1}$\\%Chk
\hline\\[-5.5mm]
\end{tabular}\end{center}
Pulsators discovered by our survey are listed with their names in bold face,
and hybrid DW\,Lyn type pulsators in italics.
$C$ gives the class of the companion, where available, and the notes in that
column mark:
$^a$\,No IR excess from {\sc 2mass} or SDSS photometry,
$^b$\,No detectable RV variations on a time-scale of hours or days.
$^c$\,The F3 companion to LM\,Dra is separated by 3.7'', but it is unclear
whether the subdwarf itself is single.
$P_\mathrm{range}$ gives the range of periods
observed, excluding marginal detections, harmonics and $g$-modes,
$A_\mathrm{max}$ gives the highest amplitude reported,
and $N_P$ the number of independent periods reported.
The keys to the references in the last column are given on the next page.
\end{table*}

\subsection{Summary}

The four new pulsators presented in this paper certainly deserve
more attention as well.
Confirming the pulsations in \object{PG\,1033+201} would be a priority,
but an intensive campaign would probably be required to disentangle its
multiperiodic nature.
The pulsations observed in \object{HE\,2151--1001}
are barely above our significance threshold and need to be confirmed.
\object{HS\,2125+1105} is clearly a stable pulsator, as the same oscillation
period was detected in six sequences spanning four years, but we would suspect
more frequencies to be discovered with more dedicated follow-up.
\object{HE\,1450--0957} is also a convincing case where residual power
in the amplitude spectrum between 6 and 10\,mHz indicates the presence of unresolved
pulsations that could be revealed with dedicated follow-up.
The existence of pulsations in this star was recently confirmed by
observations from South Africa (C.~Koen, priv.~comm.).

In Table~\ref{tbl:stat} we compare the fraction of pulsators detected
in each of the different samples we have drawn our candidates from.
A fraction of pulsators between 5 and 12\,\%\ is clear when targets
are selected based on spectroscopic temperature estimates. 
We estimate that it should be possible to obtain a pulsation fraction
of about 10\,\%, when starting with an undepleted sample and reaching
a noise level below 1\,mma. This could probably be increased by
focusing only on the hot end of the EHB, between 31 and 36\,kK,
at the cost of sacrificing discoveries of the rarer pulsators at the
cool end of the instability region.

That the colour selected sample has a lower success fraction than the
samples based on spectroscopy is not surprising, but it is surprising
that there are so few pulsators in the sample from \citet{saffer94}.
The total number of sdB
stars in that sample hotter than 28\,kK are 44, of which we observed 16
(plus another eight that were outside the borders of the instability region).
Excluding the known $g$-mode pulsators on the cool end of that sample,
we have two short period sdBVs in 41 stars; \object{QQ\,Vir} found by us and
\object{V1078\,Her} found by \citet{bonanno03}. A large number of stars from
this sample were surveyed by \citet{billeres02}, and the remaining
stars have been checked by other observers (but no limits have
been published), so we do not expect more clear pulsators in this
sample, although some low amplitude variables may have been missed. 
% Thus, the total fraction of short period pulsators for the
% \citet{saffer94} sample would appear to be only $\sim$\,5\,\%\,
% which is lower than our detection rate in the other spectroscopically
% selected samples.
Similarly, the number of sdB stars in the instability region in the
sample of \citet{maxted01} is 34. We found no pulsators in this sample,
but it has been heavily exploited by other groups; 
\citet{billeres02} surveyed 14 of them and found one:
\object{UY\,Sex}. When combining information from this and other
unpublished surveys, we can only find a single star that has yet to
be surveyed. Thus, the best fraction from the \citet{maxted01}
sample we can infer is 3\,\%, which is lower than
what we have found in the HS, {\sc spy} and SDSS samples, but with
such small samples this is barely significant.
% Note also that 6 non-pulsators are common between these two samples.

\subsection{The big picture}
We have compiled all the known short period sdB pulsators in
Table~\ref{tbl:sdbvars}, using data from the literature and added
the results from this paper. We included preliminary physical parameters
on two new pulsators published by \citet{kilkenny09}, based on recent
low-resolution spectroscopy from {\sc not/alfosc}
(3\,250\,--\,6\,150\,\AA\ at $\sim$4.5\,\AA\ resolution),
using the same atmosphere models as
in our other determinations. A few stars still lack temperature
and gravity estimates from spectroscopy, but we have supplied 
values from our own {\sc alfosc} spectroscopy where available.
The table may also not be entirely
complete. \object{PG\,0856+121}, for which \citet{piccioni00} reported what
they termed `a pulsation episode', was dropped from the table as
repeated attempts (Reed, priv.~comm) have failed to confirm any pulsations.
Such brief pulsation episodes may be common in sdB stars, but they are hard
to catch and characterise.

The first sdB pulsator candidate in a globular cluster was reported by
\citet{randall09a}, but while the single clear period at 114\,s makes
this star a very likely sdBV, spectroscopic confirmation remains to
be done.
The same is the case for the pulsator candidate reported by \citet{silvotti09},
which was found in the Kepler satellite field-of-view, with a 
likely pulsation period of 125\,s found at the 3.5$\sigma$ level.

We note that the number of pulsators in Table~\ref{tbl:sdbvars} with
F--K companions are 18 of 49 = 37\,\%, which is compatible with the fraction
of spectroscopic composites we found in the SDSS as a whole.
Pulsators also occur in
systems with close white dwarf or M-dwarf companions, which means that
pulsations appear independently of whether the subdwarf was formed through
stable Roche-lobe overflow or via common envelope ejection. The stability
of the pulsations for stars such as \object{V391\,Peg} has revealed
that some sdBs are definitely single, which implies that a
formation channel must exist that produces single sdB stars that are
otherwise indistinguishable from those produced via the channels that
produces wide sdB+F--K or short period binaries.

In Fig.~\ref{fig:tpg}a we show the location in the (\teff,\,\logg) plane
of the pulsators from our sample (blue) together with pulsators from the
literature (red), as listed in Table~\ref{tbl:sdbvars},
with symbol size proportional to $A_\mathrm{max}$.
As our sample stars are fairly evenly spaced along the canonical EHB strip,
it is immediately
clear that pulsations are more common on the hot side than on the cool
side, but the amplitudes are often much higher in the cool, low gravity region.
The figure does not really reproduce the gap around \teff\,$\sim$\,30\,kK
between the main group
of pulsators on the hot end of the EHB and the cool group of $g$-mode
and hybrid pulsators seen in Fig.~3 of \cite{ostensen09}, where
the spectroscopy was obtained from the independent BG survey
\citep{green08}. This may be due to the fact that diverse methods
and model grids are used in the temperature determinations collected from
the literature, and this can produce scatter as high as 2\,kK in
\teff\ and 0.2\,dex in \logg.
If we plot only the stars for which we ourselves have obtained the fits on a
consistent model grid, the gap is indeed present.

\let\thefootnote\relax\footnotetext{
References to Table~\ref{tbl:sdbvars}. Discovery papers:
$^1$\,\citet{kilkenny97}, $^2$\,\citet{koen97},
$^3$\,\citet{stobie97}, $^4$\,\citet{odonoghue97},
$^5$\,\citet{billeres97,odonoghue98}, $^6$\,\citet{kilkenny98_pg1336},
$^7$\,\citet{billeres98,koen98_kpd2109}, $^8$\,\citet{koen98_pg1605},
$^9$\,\citet{koen98_f48}, $^{10}$\,\citet{koen99a}, $^{11}$\,\citet{koen99b},
$^{12}$ \citet{kilkenny99},
$^{13}$\,\citet{billeres00}, $^{14}$\,\citetalias{silvotti00},
$^{15}$\,\citetalias{ostensen01a}, $^{16}$\,\citetalias{ostensen01b},
$^{17}$\,\citet{brassard01}, $^{18}$\,\citetalias{silvotti02a},
$^{19}$\,\citet{dreizler02}, $^{20}$\,\citet{bonanno03},
$^{21}$\,\citet{koen04}, $^{22}$\,\citetalias{solheim04}; \citet{aerts06},
$^{23}$\,\citet{oreiro04}, $^{24}$\,\citetalias{solheim06},
$^{25}$\,\citet{kilkenny06}, $^{26}$\,\citet{ramsay06},
$^{27}$\,\citetalias{oreiro07}, $^{28}$\,\citetalias{oreiro09},
$^{29}$ \citet{kilkenny09}, $^{30}$\,\citet{barlow09},
$^{31}$\,\citet{barlow10},
$^{32}$\,\citet{koen07}, {Kilkenny (priv.comm)},
$^\dag$ This paper.
Supplementary data:
$^{a1}$\,\citet{allard94},
$^{b1}$\,\citet{baran10}, $^{c1}$\,\citet{charpinet05}, $^{g1}$\,\citet{geier07},
$^{h1}$\,\citet{heber00}, $^{k1}$\,\citet{kilkenny02}, $^{k2}$\,\citet{kilkenny03},
$^{l1}$\,\citet{lisker05}, $^{o1}$\,\citet{oreiro05}, $^{o2}$\,\citet{otoole04},
$^{r1}$\,\citet{reed06}, $^{r2}$\,\citet{reed07b}, $^{r3}$\,\citet{reed07a}, $^{r4}$\,\citet{reed10},
$^{r5}$\,\citet{randall07}, $^{r6}$\,\citet{randall09b},
$^{s1}$\,\citet{silvotti02b,silvotti07}, $^{s2}$\,\citet{silvotti06},
$^{s3}$\,\citet{schuh06}, $^{t1}$\,\citet{telting04}, $^{v1}$\,\citet{maja06}, $^{v2}$\,\citet{maja07},
$^{z1}$\,\citet{zhou06}, $^\ddag$ Our determination of \teff, \logg\ and \logy.
}
The main cluster of pulsators correlates
very well with the initial instability strip predicted by \citet{charpi97},
with only a few rare pulsators appearing to be anomalous.
While the sample presented here contains no stars hotter than 39\,kK,
we have observed quite a few such stars without finding any with
significant variability. But these are classified as
sdO or He-sdO stars, and will be presented in a future paper.
Candidates in that sample were selected from very diverse surveys
and ended up having a wide range of physical parameters,
as stars matching the properties of the only known sdO pulsator,
\object{J16007+0748} \citep{woudt06} are extremely rare, preventing
an effective preselection such as we made for the sdB sample.

Figure \ref{fig:tpg}b shows the surface gravities plotted against the
range of pulsation periods listed in Table~\ref{tbl:sdbvars}.
As the \logg\ values are derived with many different methods, the systematic
errors can be substantially larger than $\pm$0.1 dex. Still, the expected
relationship between pulsation periods and surface gravity seems to hold
up quite well \citep[see][]{koen99a}.
The relationship appears to break down for objects with surface
gravities higher than 5.8\,dex, but this may not be a real effect. Evolutionary
tracks for EHB stars with {\em any} envelope thick enough to sustain
pulsations can hardly reach higher surface gravities than about 5.9\,dex
(note the lower end of the canonical ZAEHB in Fig.~\ref{fig:tpg}).
On the other hand,
it is well known that the very high rotation rates in binary systems,
high metallicity, as well as the pulsations themselves,
give rise to broadening of atmospheric lines.
If these broadening effects are not accounted for in the models used for
estimating the atmospheric parameters, the fitting procedure may converge
towards a too high gravity to compensate.

\section{Conclusions and outlook}

The interest in sdB pulsators can only increase in the future, when
satellite missions such as Kepler \citep{kepler}
start to find numerous new sdBVs and observe them with
unprecedented frequency resolution and at lower pulsation amplitudes
than have ever been achieved from the ground.
As we enter the age of space based asteroseismology for the faint subdwarf
stars, we can expect our understanding of the incidence of pulsations and
their amplitude variability over time and across the instability region
to improve, as
{\sc most} \citep{most} and CoRoT \citep{corot} have done for brighter
pulsating stars.
Meanwhile, a new generation of ground based instruments are coming on-line,
capable of high temporal resolution and simultaneous
multi-colour photometry, by splitting
the light with dichroics and reading out several CCDs simultaneously.
{\sc ultracam} has pioneered this technology, permitting reliable amplitude
ratios to be measured and used directly for mode identification.

As theoretical models of the interior structure of these stars become
sophisticated enough to accurately predict the pulsation frequencies
\citep{haili08,haili09},
new high-precision photometric observations will help us
distinguish between the possible formation scenarios. When this happens we
will finally be in position to unravel the complete evolutionary history of
the hot subdwarf stars.

\begin{acknowledgements}
The authors thank the staff at the {\sc not} for excellent support
over a decade of observations.

The time-series data presented here have been taken using \alfosc, which
is owned by the Instituto de Astrofisica de Andalucia (IAA) and operated
at the Nordic Optical Telescope under agreement between IAA and the
NBIfAFG of the Astronomical Observatory of Copenhagen. 

The research leading to these results has received funding from the European
Research Council under the European Community's Seventh Framework Programme
(FP7/2007--2013)/ERC grant agreement N$^{\underline{\mathrm o}}$\,227224
({\sc prosperity}), as well as from the Research Council of K.U.Leuven grant
agreement GOA/2008/04.

CRL acknowledges an {\em \'Angeles Alvari\~no} contract of
the regional government {\em Xunta de Galicia}.

\end{acknowledgements}
\bibliographystyle{aa}
\bibliography{biblio}

\end{document}